\documentclass[showpacs,amsmath,amssymb,aps,preprint,nofootinbib]{revtex4-1}

\usepackage{graphicx}
\usepackage{dcolumn}
\usepackage{color}
\usepackage{float}

\usepackage{hyperref}
\hypersetup{%
colorlinks=true,%
linkcolor=blue,
citecolor=blue,
urlcolor=magenta
}
\newcommand{\sumint}{\mbox{$\sum$}\kern-2.7ex\int}

\def\gtsim{\mathrel{\hbox{\raise0.2ex
\hbox{$>$}\kern-0.75em\raise-0.9ex\hbox{$\sim$}}}}
\def\ltsim{\mathrel{\hbox{\raise0.2ex
\hbox{$<$}\kern-0.75em\raise-0.9ex\hbox{$\sim$}}}}

\begin{document}


\title{Refined renormalization group improvement for thermally resummed effective potential}

\author{Koichi Funakubo$^{1}$}
\email{funakubo@cc.saga-u.ac.jp}
\author{Eibun Senaha$^{2,3}$}
\email{eibunsenaha@vlu.edu.vn (corresponding author)}
\affiliation{$^1$Department of Physics, Saga University,
Saga 840-8502 Japan}
\affiliation{$^2$Subatomic Physics Research Group, Science and Technology Advanced Institute, Van Lang University, Ho Chi Minh City, Vietnam}
\affiliation{$^3$Faculty of Applied Technology, School of Technology, Van Lang University, Ho Chi Minh City, Vietnam}

\date{\today}

\begin{abstract}
We newly develop a renormalization group (RG) improvement for thermally resummed effective potentials. 
In this method, $\beta$-functions are consistently defined in resummed perturbation theories, so that order-by-order RG invariance is not spoiled after thermal resummation. With this improvement, scale dependences of phase transition quantities such as a critical temperature, which are known to be notoriously large at the one-loop order, are greatly reduced compared to calculations with the conventional $\overline{\text{MS}}$ scheme. 
By taking advantage of the RG invariance, we also devise a resummation method that can incorporate potentially harmful large logarithmic terms and temperature-dependent power corrections in a generic form. We point out that a resummed one-loop effective potential refined by the method can give results that agree with those obtained by resummed two-loop effective potentials within errors.

\end{abstract}


\maketitle


\section{Introduction}

Investigating phase transitions in the early Universe is expected to shed light on new physics searches in particle physics and cosmology. 
Much attention has been drawn to gravitational wave generations from first-order phase transitions, which could provide useful information on high energy physics that cannot be obtained by terrestrial experiments. Furthermore, if electroweak phase transition (EWPT) is first order, a cosmic baryon asymmetry can be explained by electroweak baryogenesis (EWBG) mechanism~\cite{Rubakov:1996vz,*Funakubo:1996dw,*Riotto:1998bt,*Trodden:1998ym,*Bernreuther:2002uj,*Cline:2006ts,*Morrissey:2012db,*Konstandin:2013caa,*Senaha:2020mop}. 

While nonperturbative approaches such as lattice calculations would be robust, perturbative treatments are still useful for probing vast parameter space in new physics models because of their lower costs.
One of the vexing problems at finite temperature is infrared divergences originating from a zero Matsubara frequency mode, which could spoil the validity of perturbative expansions even for small coupling constants at high temperature~\cite{Dolan:1973qd,Linde:1980ts}.
It is standard practice to reorganize the perturbative expansion to incorporate the dominant temperature corrections into the unperturbed part, which is referred to \textit{thermal resummation}~\cite{Parwani:1991gq,Carrington:1991hz,Arnold:1992rz}. 
One-loop effective potentials with resummation schemes in Refs.~\cite{Parwani:1991gq,Carrington:1991hz,Arnold:1992rz} have been mostly employed in studies of EWPT (for other approaches, see, e.g., Refs.~\cite{Gould:2021oba,*Croon:2020cgk,*Schicho:2022wty,Athron:2022jyi}).

In perturbative analyses of EWPT, a renormalization scheme dependence inevitably comes into calculations, and the magnitude of which implies impacts of higher-order terms that are missing in the calculations. If the dependence is too large to make quantitative studies reliable, a renormalization group equation (RGE) can be used to improve the calculations~\cite{Coleman:1973jx,Kastening:1991gv,Bando:1992np,*Bando:1992wy,Ford:1992mv}.
This can be done by replacing parameters appearing in the effective potential with corresponding running parameters derived from $\beta$-functions which are perturbatively defined at some fixed order. 
One should note that the derivation of the $\beta$-functions follows from the scale independence of bare parameters together with a specific renormalization scheme such as a $\overline{\text{MS}}$ scheme~\cite{Machacek:1983tz,*Machacek:1983fi,*Machacek:1984zw,Luo:2002ey,*Luo:2002ti}.
Once the effective potential is made scale independent at some order, one can incorporate a series of higher-order terms utilizing its scale invariance.  
As demonstrated in Refs.~\cite{Kastening:1991gv,Bando:1992np,*Bando:1992wy} at zero temperature, a $\ell$-loop effective potential with $(\ell+1)$-loop $\beta$-functions can resum up to $\ell$th-to-leading logarithmic terms.
At nonzero temperature, however, such an RG improvement of the effective potential would not be straightforward due to the aforementioned thermal resummation. Unlike the zero temperature case, the order-by-order RG invariance is lost, and higher-order terms are required to recover the RG invariance up to a certain order in coupling constants. For example, the RG invariance of the resummed one-loop effective potential requires a part of two-loop effective potentials. Explicit calculations using a high-temperature expansion can be found in Ref.~\cite{Arnold:1992rz} (for a recent study, see Ref.~\cite{Gould:2021oba}). 
Another difference from the zero temperature is that in addition to the potentially large logarithmic terms, temperature-dependent power corrections could also be sizable at higher temperatures, as described above. Thus, the commonly used log-resummation scheme is not always appropriate. 
In light of this situation, the main issues to be clarified are as follows:
\begin{itemize}
\item How do we construct an order-by-order RG invariant effective potential at finite temperature?  
\item How do we incorporate both logarithmic terms and temperature-dependent corrections in a general manner?
\end{itemize}
In our recent letter paper~\cite{Funakubo:2023cyv}, we proposed a novel RG improvement method for the resummed effective potentials to answer the above questions. In our method, $\beta$-functions are defined in the resummed perturbation theory instead of using those in the $\overline{\text{MS}}$ scheme, and as a result, the RG invariance is maintained order by order after the thermal resummation.
In addition to this, the resummation by RG is generalized to include whole loop functions that contain both logarithmic terms and thermal corrections. 
By its general form, this method is reduced to the log-resummation scheme in the zero temperature limit, while the hard thermal loop resummation in the high-temperature limit.
Due to the length limitation of the letter~\cite{Funakubo:2023cyv}, we show only a main result, and some details are omitted. 

In this paper, we fill the gap in Ref.~\cite{Funakubo:2023cyv} by giving all the details, including lengthy but useful expressions, and adding more numerical examples to clarify our method further. 
One of the main findings is that the resummed one-loop effective potential in our scheme has much less scale dependence than that in the $\overline{\text{MS}}$ scheme thanks to the order-by-order RG invariance, though an exceptional region can, in principle, be found due to an accidental cancellation between RG-noninvariant terms and truncation errors in the $\overline{\text{MS}}$ scheme.
If one considers two-loop corrections, both schemes are equally better than the one-loop result in our scheme. 
This is because the two-loop corrections cancel the dominant RG-noninvariant terms in the $\overline{\text{MS}}$ scheme. 
As a by-product of the RG invariance in our scheme, a series of higher-order terms can be incorporated into the resummed effective potentials.
In the case of a single field theory such as the $\phi^4$ theory, we can show that the resummed one-loop effective potential in our method correctly reproduces dominant two-loop corrections. 
Even in a two-scalar field theory, our numerical studies show that $v_C/T_C$ obtained by the resummed one-loop effective potential with our two-loop $\beta$-functions falls within the two-loop order scale uncertainties, where $T_C$ denotes a critical temperature and $v_C$ is a vacuum expectation value (VEV) at $T_C$.
Therefore, our RG-improved effective potential would be particularly useful when the complete two-loop effective potential is unavailable. 

The paper is organized as follows. In Sec.~\ref{sec:beta}, $\beta$-functions of masses and couplings and $\gamma$-functions of fields are generally derived by employing the dimensional regularization. 
In Sec.~\ref{sec:phi4}, as the first application, we demonstrate the RG invariance of the effective potentials up to the two-loop order in the $\phi^4$ theory and make a comparison between the $\overline{\text{MS}}$ and our schemes analytically and numerically.
We also present how to incorporate higher-order terms based on the RG invariance at some fixed order.
An application of our method to the $\phi^4$ theory with an additional real scalar field is conducted in Sec.~\ref{sec:Exphi4}.
The numerical results of first-order phase transitions are presented in this section. Sec.~\ref{sec:conclusion} is devoted to the conclusion.
Some detailed expressions are given in Appendices. 

\section{$\beta$-functions in the resummed theory}\label{sec:beta}
Let us collectively denote arbitrary fields and couplings as $\phi_i(x)$ and $g_k$
and boson and fermion masses as $m_a^2$, and $M_\alpha$, respectively, and a vacuum energy is denoted as $\Omega$.
We use dimensional regularization in which the spacetime dimension is analytically continued to the $d=4-\epsilon$ dimension~\cite{tHooft:1972tcz}. 
In this case, the mass dimensions of the bare couplings $g_{Bk}^{}$ become $\sigma_k\epsilon$, where $\sigma_k=1$ for scalar quartic couplings and $\sigma_k=1/2$ for gauge and Yukawa couplings, respectively, while that of the bare vacuum energy $\Omega_B$ is $d$. 
Before discussing our scheme, we begin by deriving $\beta$-functions in mass-independent regularization schemes such as $\text{MS}$ and $\overline{\text{MS}}$~\cite{Machacek:1983tz,*Machacek:1983fi,*Machacek:1984zw,Luo:2002ey,*Luo:2002ti}. The bare parameters are decomposed into the renormalized parts and $\epsilon$ poles:
\begin{align}
g_{Bk}^{}\mu^{-\sigma_k\epsilon} & = g_k +\sum_{n=1}^\infty\frac{a_k^{(n)}(g)}{\epsilon^n}, \\
m_{Ba}^2 & = \left(\delta_{ab}+\sum_{n=1}^\infty\frac{b_{ab}^{(n)}(g)}{\epsilon^n} \right)m_b^2,\label{mB} \\
M_{B\alpha} & = \left(\delta_{ab}+\sum_{n=1}^\infty\frac{B_{ab}^{(n)}(g)}{\epsilon^n} \right)M_\beta, \\
Z_{ij} & = \delta_{ij}+\sum_{n=1}^\infty\frac{c_{ij}^{(n)}(g)}{\epsilon^n}, \\
\Omega_B\mu^\epsilon  & = \Omega+\sum_{n=1}^\infty\frac{\omega_n(g)}{\epsilon^n}.
\end{align}
From those expressions, one can find the $\beta$-functions of each parameter as
\begin{align}
\beta_k &= \lim_{\epsilon\to0} \mu\frac{d g_k}{d\mu} = -\sigma_ka_k^{(1)}+\sum_\ell a_{k,\ell}^{(1)} \sigma_\ell g_\ell, \\
m_a^2\beta_{m_a^2} &= \lim_{\epsilon\to0} \mu\frac{d m_a^2}{d\mu} = \sum_{k,b}b_{ab,k}^{(1)}\sigma_kg_km_b^2, \\
M_\alpha\beta_{M_\alpha} &= \lim_{\epsilon\to0} \mu\frac{d M_\alpha}{d\mu} = \sum_{k,\beta}B_{\alpha\beta,k}^{(1)}\sigma_kg_kM_\beta, \\
\gamma_{ij} &=  \lim_{\epsilon\to0}\mu\frac{d Z_{ij}}{d\mu}  = -\frac{1}{2}\sum_kc_{ij,k}^{(1)}\sigma_kg_k, \\
 \beta_\Omega & =\lim_{\epsilon\to0} \mu\frac{d \Omega}{d\mu} = \omega_1,\label{betaOmega}
\end{align}
where $a_{k,l}^{(1)}=da_k^{(1)}/dg_\ell$, $b_{ab,k}^{(1)}=db_{ab}^{(1)}/dg_k$, $B_{\alpha\beta,k}^{(1)}=db_{\alpha\beta}^{(1)}/dg_k$, and $c_{ij,k}^{(1)}=dc_{ij}^{(1)}/dg_k$. 
For illustrative purposes, we focus exclusively on scalar theories throughout this paper. 

Following the work of Parwani~\cite{Parwani:1991gq}, we reorganize the Lagrangian as
\begin{align}
\mathcal{L}_B = \mathcal{L}_R+\mathcal{L}_{\text{CT}} = \left[\mathcal{L}_R-\frac{1}{2}\Sigma_a(T)\phi_a^2\right] +\left[\mathcal{L}_{\text{CT}} +\frac{1}{2}\Sigma_a(T)\phi_a^2\right],
\label{LB}
\end{align}
where $\Sigma_a(T)$ are dominant thermal corrections to the masses of the scalar fields $\phi_a$. 
$\Sigma_a(T)$ is supposed to be obtained by gap equations or other methods in advance.
At the leading order, one would have $\Sigma_a(T)=\mathcal{O}(g_i T^2)$, where $g_i$ are scalar quartic couplings. 
Though this reorganization does not change the bare Lagrangian, 
$\Sigma_a(T)$ appearing in the first square brackets are regarded as the zeroth order in this new perturbation theory, while those in the second ones are part of the counterterm (CT) which are one-order higher in this perturbative expansion (called \textit{thermal counterterm} hereafter). 
In our method, the bare mass parameters of the scalar fields in resummed perturbation theory are defined as 
\begin{align}
m_{Ba}^2 & = \left(\delta_{ab}+\sum_{n=1}^\infty\frac{b_{ab}^{(n)}(g)}{\epsilon^n} \right)m_b^2+\sum_{n=1}^\infty\frac{\tilde{b}_{ab}^{(n)}(g)}{\epsilon^n}\Sigma_b(T),
\label{m2B}
\end{align}
where the last terms correspond to temperature-dependent divergences. 
Such terms must be absent in all-order calculations since the divergence structure of the theory must not be altered by the thermal resummation.
At a fixed order in the resummed perturbation theory, however, one would encounter temperature-dependent divergences, as seen in the actual effective potential calculations shown in the next section.
Even though the \textit{new} divergences are expected to be cancelled by higher-order terms, the order-by-order renormalizability would be generally unclear. On the other hand, if CTs are defined in the form of Eq.~(\ref{m2B}) at each order in the resummed perturbation theory, the renormalization would be more apparent. This is the strategy we adopt here. 

The rearrangement of the perturbative expansion seems to mess up the order-by-order RG invariance. 
While the scaling of $\Sigma_a(T)$ may be nontrivial, it should be scale independent for full-order calculations, and thus the scaling of the resummed effective potential would not be altered.
In principle, it is possible to construct $\Sigma_a(T)$ in a self-consistent way by solving a complete set of Schwinger-Dyson equations. However, from a practical standpoint, we assume that $\Sigma_a(T)$ is preset as a solution to the gap equation in a scheme different from the one we are considering here. For illustration, we have shown in Appendix~\ref{app:SigmaT} that $\Sigma_a(T)$ adopted here is scale invariant up to the two-loop level in the $\overline{\text{MS}}$ scheme. As always, there exists a residual scale dependence in perturbatively calculated $\Sigma_a(T)$.  
However, this scale dependence is a matter of precision when computing it, and in principle, we could improve it by including higher-order terms in the gap equation.
From our standpoint, this is a separate matter from the scale dependence issue of the effective potential that we will discuss below, and we do not put the two different scale dependencies together for consistency. 
The scale invariance of $\Sigma_a(T)$ allows us to choose the couplings at a particular fixed scale $g_i(\mu_\text{fixed})$ for $\Sigma_a(T)$. For the sake of simplicity, we use initial values of the RG running for $\Sigma_a(T)$ and a high-temperature approximation, as detailed in Appendix~\ref{app:SigmaT}.
In this paper, we call $d\Sigma_a(T)/d\mu=0$ \textit{consistency condition}. 
With this condition, we prove the order-by-order RG invariance of the resummed effective potentials up to the two-loop level.\footnote{The RG-invariant resummed pressure using a different method can be found in Ref.~\cite{Kneur:2015uha,*Kneur:2015moa}.}
Following the same step as in the $\overline{\text{MS}}$ scheme but with the consistency condition, it follows that
\begin{align}
m_a^2\beta_{m_a^2} =\sum_{k,b}\big(b_{ab,k}^{(1)}m_b^2+\tilde{b}_{ab,k}^{(1)}\Sigma_b\big)\sigma_kg_k.\label{beta_m2}
\end{align}
The thermal resummation also generates temperature-dependent divergences in the vacuum energy. 
However, the relation $\beta_\Omega=\omega_1$ is not altered once the consistency condition is imposed.
Furthermore, $\beta$-functions of dimensionless couplings remain the same as those in the $\overline{\text{MS}}$ scheme.

\section{$\phi^4$ theory} \label{sec:phi4}
We first consider the $\phi^4$ theory to explain our scheme and show the order-by-order RG invariance up to the two-loop levels.  
The bare Lagrangian is given by
\begin{align}
\mathcal{L}_B = \frac{1}{2}\partial_\mu \Phi_B \partial^\mu \Phi_B-V_B(\Phi_B),\quad
V_B(\Phi_B)=\Omega_B-\frac{\nu_B^2}{2}\Phi^2+\frac{\lambda_B}{4!}\Phi_B^4.\label{bLag_phi4}
\end{align}
As shown below, the vacuum energy $\Omega$ is also needed to show the RG invariance of the effective potentials. 
We decompose $\mathcal{L}_B$ into the renormalized Lagrangian ($\mathcal{L}_R$) and CT ($\mathcal{L}_{\text{CT}}$), and subtract and add a dominant thermal correction $\Sigma(T)$ in each part. The explicit form of the resummed Lagrangian is given in Appendix~\ref{app:phi4}.

We derive the effective potentials up to the two-loop level in this resummed perturbation theory. 
Let us denote the classical background field as $\varphi$. The tree-level effective potential is 
\begin{align}
V_0(\varphi)&=\Omega+\frac{1}{2}\left(-\nu^2+\Sigma(T)\right)\varphi^2+\frac{\lambda\mu^\epsilon}{4!}\varphi^4,
\end{align}
where $\Sigma(T)$ must be regarded as the zeroth-order term. The field-dependent mass is given by
\begin{align}
M^2 = \frac{\partial^2 V_0}{\partial \varphi^2}=m^2 + \Sigma(T),
 \end{align}
with $m^2 = -\nu^2+\lambda\mu^\epsilon\varphi^2/2$. Consequently, the resummed one-loop effective potential takes the form
\begin{align}
\mu^\epsilon V_1(\varphi)&= \frac{M^4}{4(16\pi^2)}\left(-\frac{2}{\epsilon}+\ln\frac{M^2}{\bar{\mu}^2}-\frac{3}{2}+\mathcal{O}(\epsilon)\right),\label{V1phi4}
\end{align}
where $\bar{\mu} = \sqrt{4\pi e^{-\gamma_E^{}}} \mu \simeq 2.66 \mu$ with $\gamma_E^{}$ being the Euler constant. As mentioned in Sec.~\ref{sec:beta}, the temperature-dependent divergence appears in the fixed-order calculation. 
In our renormalization scheme, the whole divergences in Eq.~(\ref{V1phi4}) are removed  by CTs defined in Eqs.~(\ref{CTOmega})-(\ref{CTlam}), leading to
\begin{align}
\delta^{(1)}\Omega = \frac{1}{\epsilon}\frac{(\nu^2-\Sigma)^2}{32\pi^2},\quad
\delta^{(1)}\nu^2 = \frac{1}{\epsilon}\frac{\lambda(\nu^2-\Sigma)}{16\pi^2}, \quad
\delta^{(1)}\lambda = \frac{1}{\epsilon}\frac{3\lambda^2}{16\pi^2}.
\label{1LCT}
\end{align}
Therefore, CTs of the dimensionful parameters are modified by the thermal resummation. 
With those CTs, the bare mass parameters $\Omega_B$ and $\nu_B$ are expressed as
\begin{align}
\Omega_B\mu^\epsilon &= \Omega+\delta^{(1)}\Omega = \Omega+\frac{1}{\epsilon}\frac{(\nu^2-\Sigma)^2}{32\pi^2}, \\
\nu_B^2 &= Z_\Phi^{-1}(\nu^2+\delta^{(1)} \nu^2) = 
\nu^2\left(1+\frac{1}{\epsilon}\frac{\lambda}{16\pi^2}\right)-\Sigma\left(\frac{1}{\epsilon}\frac{\lambda}{16\pi^2}\right)\label{nu2B_1L},
\end{align}
where $Z_\Phi=1$ at the one-loop level. From our $\beta$-function formulas (\ref{betaOmega}) and (\ref{beta_m2}), it follows that (for the derivation, see Appendix~\ref{app:phi4})
\begin{align}
\beta_\Omega^{(1)} & = \frac{(\nu^2-\Sigma)^2}{32\pi^2},\\
\nu^2\beta_{\nu^2}^{(1)} &= \frac{\lambda(\nu^2-\Sigma)}{16\pi^2}.\label{beta1_nu2}
\end{align}
In the limit of $\Sigma=0$, our $\beta$-functions are reduced to those in the $\overline{\text{MS}}$ scheme.
Therefore, the difference between the two schemes could be sizable when $\Sigma$ is comparable to $\nu^2$. 
If one uses CTs in the $\overline{\text{MS}}$ scheme, the temperature-dependent divergences would remain at this order. As pointed out in Ref.~\cite{Karsch:1997gj} (see also Ref.~\cite{Laine:2017hdk}), higher-order loop corrections are needed to cancel such divergences.\footnote{
One could consider a resummation method shown in Ref.~\cite{Banerjee:1991fu,Chiku:1998kd}, in which the bare Lagrangian is decomposed into
\begin{align}
\mathcal{L}_B = \mathcal{L}_R+\mathcal{L}_{\text{CT}} &= \left[\frac{1}{2}(\partial_\mu \Phi)^2+\frac{1}{2}M^2\Phi^2-\frac{\lambda}{4!}\Phi^4 +\frac{1}{2}\Sigma\Phi^2 \right] \nonumber\\
&\quad+\left[\frac{A}{2}(\partial_\mu \Phi)^2+\frac{B}{2}(M^2-\Sigma)\Phi^2-C\frac{\lambda}{4!}\Phi^4+D(M^2-\Sigma)^2 \right],
\end{align}
where $M^2=\nu^2-\Sigma$ and $A$, $B$, $C$, and $D$ are CTs in the $\overline{\text{MS}}$ scheme at zero temperature. Orders (denoted as $\delta$) of $M^2$ and $\Sigma$ in the resummed perturbation theory are regarded as $M^2 = \mathcal{O}(\delta^0)$ and $\Sigma = \mathcal{O}(\delta)$. With this order counting, the one-loop CTs for the mass and vacuum energy are reduced to
$\frac{B}{2}M^2\Phi^2+DM^4$, which are essentially the same as our CTs, and the order-by-order renormalization works.
}

As alluded to above, the dimensionless quantities are not affected by the thermal resummation considered here, and thus 
the $\beta$-function of $\lambda$ and $\gamma$-function are the same as those in the ordinary $\overline{\text{MS}}$ scheme, i.e., 
\begin{align}
\beta_\lambda^{(1)} =  \frac{3\lambda^2}{16\pi^2},\quad \gamma_\Phi^{(1)} = 0.
\end{align}
The resummed one-loop effective potential after subtracting the divergences amounts to
\begin{align}
V_{\text{eff}}(\varphi) = V_0(\varphi)+V_1(\varphi),
\label{RVeff_1L}
\end{align}
where
\begin{align}
V_0(\varphi)&=\Omega+\frac{1}{2}\left(-\nu^2+\Sigma(T)\right)\varphi^2+\frac{\lambda}{4!}\varphi^4, \label{RV0} \\
V_1(\varphi) &= \frac{M^4}{4(16\pi^2)}
\left(
	\ln\frac{M^2}{\bar{\mu}^2}-\frac{3}{2}
\right)
+\frac{T^4}{2\pi^2}I_B(A^2)-\frac{1}{2}\Sigma(T)\varphi^2,\label{RV1}
\end{align}
with $A^2=M^2/T^2$ and the thermal function $I_B(A^2)$ is defined as
\begin{align}
I_B(A^2) &= \int_0^\infty dx~x^2\ln\left(1-e^{-\sqrt{x^2+A^2}}\right) \\
&\simeq -\frac{\pi^4}{45}+\frac{\pi^2}{12}A^2-\frac{\pi}{6}(A^2)^{3/2}
	-\frac{A^4}{32}\bigg(\ln\frac{A^2}{\alpha_B}-\frac{3}{2}\bigg),
\label{IBHTE}
\end{align}
where the high-temperature expansion (HTE) is used in the second line, and $\ln\alpha_B=2\ln4\pi-2\gamma_E\simeq 3.91$.
The last term in $V_1(\varphi)$ comes from the thermal CT which avoids the double counting of $\Sigma(T)\varphi^2/2$.

Now, we move on to the two-loop analysis.  As is the one-loop case, all the divergences appearing in the two-loop effective potential are removed by CTs defined in Eqs.~(\ref{CTOmega})-(\ref{CTZPhi}). Correspondingly, the $\beta$-functions of the theory parameters in our scheme are found to be
\begin{align}
\gamma_\Phi^{(2)} &= \frac{\lambda^2}{12(16\pi^2)^2}, \\
\beta_\Omega^{(2)} & = \frac{(\nu^2-\Sigma)\Sigma}{16\pi^2}, \\
\nu^2\beta^{(2)}_{\nu^2} &= \frac{\lambda^2(-\nu^2+\Sigma)}{(16\pi^2)^2}+\frac{\lambda\Sigma}{16\pi^2}+2\nu^2\gamma_\Phi^{(2)} \nonumber\\
 &= \frac{\lambda^2}{(16\pi^2)^2}\left(-\frac{5\nu^2}{6}+\Sigma\right)+\frac{\lambda\Sigma}{16\pi^2},
\label{beta2_nu2} \\
\beta_\lambda^{(2)} &=  -\frac{6\lambda^3}{(16\pi^2)^2}+4\lambda\gamma_\Phi^{(2)}=\frac{1}{(16\pi^2)^2}\left(-\frac{17\lambda^3}{3}\right).
\end{align}
Similarly to the one-loop order, only $\beta$-functions of the dimensionful parameters are modified by the thermal resummation. 
One can see that there exists $\lambda\Sigma/(16\pi^2\nu^2)$ in $\beta_{\nu^2}^{(2)}$ which is the same as the temperature-dependent term in Eq.~(\ref{beta1_nu2}) but the opposite sign. At first sight, they appear to be canceled out in $\beta_{\nu^2} = \beta_{\nu^2}^{(1)}+\beta_{\nu^2}^{(2)}$. 
As shown in the RG invariance of the effective potential using HTE, however, one has to regard $\lambda\Sigma/(16\pi^2\nu^2)$ in $\beta_{\nu^2}^{(2)}$ as one-order higher correction than that in $\beta_{\nu^2}^{(1)}$, implying that $\lambda$ appearing in the former is one-order lower than that in the latter (see also Eq.~(\ref{nu2_2L}) below). 
We also note that $\beta_\Omega^{(2)}$ is nonzero due to the thermal correction, which is another difference from the $\overline{\text{MS}}$ scheme.

After removing all the divergences by CTs, the two-loop corrections to the resummed effective potential are cast into the form
\begin{align}
V_2(\varphi) &= 
\frac{\lambda}{8}\bar{I}^2(M)-\frac{\lambda^2\varphi^2}{12}\tilde{H}(M)-\frac{1}{2}\Sigma(T)\bar{I}(M),
\label{RV2}
\end{align}
where the loop functions $\bar{I}(M)$ and $\tilde{H}(M)$ are defined in Eqs.~(\ref{barI}) and (\ref{tilH}), respectively.
The last term in Eq.~(\ref{RV2}) corresponds to the thermal CT at this order, and by which the double counting of $\Sigma(T)$ corrections and linear-like terms in $\varphi$ such as $\mathcal{O}((M^2)^{1/2}T^3)$ are avoided~\cite{Arnold:1992rz}. 

\subsection{RG invariance of the thermally resummed effective potential}\label{subsec:RGinv}
Now that we have obtained the renormalized effective potentials and $\beta$-functions in our scheme at one- and two-loop orders, we show their RG invariance one by one.  
The effective potential satisfies~\cite{Coleman:1973jx,Kastening:1991gv,Bando:1992np,*Bando:1992wy,Ford:1992mv}
\begin{align}
0 = \mu\frac{ dV_{\text{eff}}}{d\mu} = 
\left[
\mu\frac{\partial}{\partial \mu}+\nu^2\beta_{\nu^2}\frac{\partial}{\partial \nu^2}+\beta_\lambda\frac{\partial}{\partial\lambda}-\gamma_\Phi^{}\varphi\frac{\partial}{\partial \varphi}+\beta_\Omega\frac{\partial}{\partial\Omega}
\right]V_{\text{eff}}\equiv \mathcal{D}V_{\text{eff}}. \label{RGinv}
\end{align}
We first show the RG invariance of the resummed one-loop effective potential. Applying the derivative operator $\mathcal{D}$ to the potential (\ref{RVeff_1L}), one gets
\begin{align}
\mathcal{D}V_0|_{\text{one-loop}} & = \beta^{(1)}_\Omega-\frac{\nu^2}{2}\beta_{\nu^2}^{(1)}\varphi^2+\frac{1}{4!}\beta_\lambda^{(1)} \varphi^4 
=\frac{M^4}{32\pi^2}, \label{DV0_1L}\\
\mathcal{D}V_1|_{\text{one-loop}} &= \mu \frac{\partial V_1}{\partial \mu}=-\frac{M^4}{32\pi^2},
\label{DV1_1L}
\end{align}
where the consistency condition $\mathcal{D}\Sigma=0$ is used. Thus, one obtains $\mathcal{D}(V_0+V_1)|_{\text{one-loop}}=0$.
We note that this invariance is due to the modified $\beta$-functions. In order words, the $\overline{\text{MS}}$ $\beta$-functions cannot maintain the RG invariance at this order. Let us consider the errors of both schemes. In our scheme, we have the truncation error which starts from the two-loop order, $\mathcal{O}(1/(16\pi^2)^2)$. In the $\overline{\text{MS}}$ scheme, on the other hand, an additional error comes from the RG-noinvariant terms, which are found to be   
\begin{align}
\mathcal{D} (V_0+V_1)_{\text{one-loop}}^{\overline{\text{MS}}} &= \frac{-(2m^2+\Sigma)\Sigma}{32\pi^2}
+\mathcal{O}\left(\frac{1}{(16\pi^2)^2)}\right) \nonumber\\
& \to \frac{-\lambda\varphi^2\Sigma}{32\pi^2}+\mathcal{O}\left(\frac{1}{(16\pi^2)^2)}\right),
\label{DV1Lms}
\end{align}
where $\varphi$-independent terms are dropped after the right arrow assuming $\Sigma=\lambda T^2/24$.
Therefore, in the $\overline{\text{MS}}$ scheme, there could be a cancellation between the two different errors depending on model parameters. We will exemplify such a case below.
However, we emphasize that the less-scale dependence is merely accidental and has no theoretical reasoning. 

It would also be instructive to see the above RG invariance using HTE. 
This demonstration focuses exclusively on the $\varphi$-dependent terms and omits the vacuum energy $\Omega$.
Using HTE of $I_B$ given in Eq.~(\ref{IBHTE}), the resummed one-loop effective potential (\ref{RVeff_1L}) up to $\mathcal{O}(\varphi^4)$ is approximated as
\begin{align}
V_{\text{eff}}^{\text{HTE}}(\varphi)
&=V_0(\varphi)+V_1^{\text{HTE}}(\varphi) \nonumber\\
&\simeq \frac{1}{2}
\bigg[
(-\nu^2+\Sigma)\left(1+\frac{\lambda}{32\pi^2}\ln\frac{T^2}{\bar{\mu}^2}\right)+\frac{\lambda(-\nu^2+\Sigma)c_B}{16\pi^2}
\bigg]\varphi^2 \nonumber\\
&\quad -\frac{T(M^2)^{3/2}}{12\pi}
+\frac{1}{4!}
\left[\lambda\left( 1+\frac{3\lambda}{32\pi^2}\ln\frac{T^2}{\bar{\mu}^2}\right)
+\frac{3\lambda^2c_B}{16\pi^2}
\right]\varphi^4,\label{V1LHTE}
\end{align}
where $c_B=(\ln\alpha_B)/2$. To make the RG invariance of $V_{\text{eff}}^{\text{HTE}}(\varphi)$ manifest, we solve $\beta_{\nu^2}^{(1)}$ and $\beta_\lambda^{(1)}$ perturbatively.
Let us denote a running parameter as $\bar{\mathcal{X}}(t)$ with $t=\ln(\bar{\mu}/\bar{\mu}_0)$, where $\bar{\mu}$ is an arbitrary scale and $\bar{\mu}_0$ is its initial value. $\bar{\mathcal{X}}(t)$ can be expanded as
\begin{align}
\bar{\mathcal{X}}(t) &= \bar{\mathcal{X}}(0) + \frac{d\bar{\mathcal{X}}(t)}{dt}\bigg|_{t=0}t+\frac{1}{2}\frac{d^2\bar{\mathcal{X}}(t)}{dt^2}\bigg|_{t=0}t^2+\cdots \nonumber\\
& = \bar{\mathcal{X}}(0) + \big(\beta_{\mathcal{X}}^{(1)} +\beta_{\mathcal{X}}^{(2)}\big)\big|_{t=0}t+\frac{1}{2}\frac{d\beta_{\mathcal{X}}^{(1)}(t)}{dt}\bigg|_{t=0}t^2+\cdots.
\label{t-expansion}
\end{align}
Using this expansion, $\bar{\nu}^2(t)$ and $\bar{\lambda}(t)$ to $\mathcal{O}(t)$ are, respectively, given by
\begin{align}
-\bar{\nu}^2(t)+\Sigma &\simeq (-\nu_0^2+\Sigma)\left(1+\frac{\lambda_0}{16\pi^2} t \right),\label{nu2_1L} \\
\bar{\lambda}(t) & \simeq \lambda_0\left(1+\frac{3\lambda_0}{16\pi^2} t \right),\label{lam_1L}
\end{align}
where $\nu_0^2=\bar{\nu}^2(t=0)$, and $\lambda_0=\bar{\lambda}(t=0)$. As noted in Sec.~\ref{sec:beta}, $\Sigma$ is given by the parameters at $t=0$.
Using those expressions, $V_{\text{eff}}^{\text{HTE}}(\varphi)$ is rewritten as
\begin{align}
V_{\text{eff}}^{\text{HTE}}(\varphi)
&\simeq \frac{1}{2}
\bigg[
-\bar{\nu}^2(T)+\Sigma+\frac{\lambda(-\nu^2+\Sigma)c_B}{16\pi^2}
\bigg]\varphi^2-\frac{T(M^2)^{3/2}}{12\pi}
+\frac{1}{4!}
\left[\bar{\lambda}(T)+\frac{3\lambda^2c_B}{16\pi^2}
\right]\varphi^4,
\end{align}
where $\bar{\nu}^2(T)$ and $\bar{\lambda}(T)$ are the running parameters evaluated at $T$ evolved from the scale $\bar{\mu}$. 
Therefore, $V_{\text{eff}}^{\text{HTE}}(\varphi)$ is manifestly RG invariant, where the explicit scale dependences of $\bar{\mu}$ are absorbed into the running parameters. This is not the case if one uses the $\beta$-functions in the $\overline{\text{MS}}$ scheme. Suppose that $\Sigma=\lambda T^2/24$, Eq.~(\ref{V1LHTE}) is cast into the form
\begin{align}
V_{\text{eff}}^{\text{HTE}}(\varphi)
&\simeq \frac{1}{2}
\bigg[
-\bar{\nu}^2(T)|_{\overline{\text{MS}}}+\frac{\lambda(-\nu^2+\Sigma)c_B}{16\pi^2}
+\frac{\lambda T^2}{24}\left(1+\frac{\lambda}{32\pi^2}\ln\frac{T^2}{\bar{\mu}^2}\right)
\bigg]\varphi^2 \nonumber\\
&\hspace{1.5cm} 
-\frac{T(M^2)^{3/2}}{12\pi}
+\frac{1}{4!}
\left[\bar{\lambda}(T)+\frac{3\lambda^2c_B}{16\pi^2}
\right]\varphi^4,
\end{align}
where $\bar{\nu}^2(T)|_{\overline{\text{MS}}}=\bar{\nu}^2(T)|_{\Sigma=0}$.
Note that the explicit $\bar{\mu}$-dependence appearing in the $\lambda T^2/24$ term of the first line cannot be absorbed into $\bar{\lambda}$ since the coefficient of $\lambda\ln(T^2/\bar{\mu}^2)/32\pi^2$ is different from the right one in Eq.~(\ref{lam_1L}), reflecting the RG noninvariance in the $\overline{\text{MS}}$ scheme. Actually, this RG-noninvariant term is also inferred from Eq.~(\ref{DV1Lms}). 
As shown below, the RG noninvariant term would become the RG-invariant form if one adds two-loop corrections~\cite{Arnold:1992rz}.

Now, we discuss the RG invariance at the two-loop level. Applying the derivative operator $\mathcal{D}$ to the resummed effective potentials (\ref{RV0}), (\ref{RV1}), and (\ref{RV2}), respectively, each contribution at the two-loop level is calculated as
\begin{align}
\mathcal{D}V_0|_{\text{two-loop}} 
&= \beta_\Omega^{(2)}-\frac{\nu^2}{2}\beta_{\nu^2}^{(2)}\varphi^2+\frac{1}{4!}\beta_\lambda^{(2)}\varphi^4+(\nu^2-\Sigma)\gamma_\Phi^{(2)}\varphi^2-\frac{1}{3!}\gamma_\Phi^{(2)}\varphi^4 \nonumber\\
& =-\frac{\lambda^2 M^2}{2(16\pi^2)^2}\varphi^2-\frac{\Sigma M^2 }{16\pi^2}-\gamma_\Phi^{(2)}\Sigma \varphi^2, \\
\mathcal{D}V_1|_{\text{two-loop}} 
&= \left[\nu^2\beta_{\nu^2}^{(1)}\frac{\partial}{\partial \nu^2}+\beta_\lambda^{(1)}\frac{\partial }{\partial \lambda}-\gamma_\Phi^{(2)}\varphi\frac{\partial }{\partial \varphi} \right]V_1 \nonumber\\
& = \frac{\lambda(M^2+\lambda \varphi^2)}{2(16\pi^2)}\bar{I}(M)+\gamma_\Phi^{(2)}\Sigma \varphi^2, \\
\mathcal{D}V_2|_{\text{two-loop}} 
& =\mu \frac{\partial V_2}{\partial \mu}
=\frac{\lambda^2M^2\varphi^2}{2(16\pi^2)^2}
-\frac{\lambda (M^2+\lambda\varphi^2)}{2(16\pi^2)}\bar{I}(M)
+\frac{\Sigma M^2}{16\pi^2}.
\end{align}
Summing up, one verifies that $\mathcal{D}(V_0+V_1+V_2)|_{\text{two-loop}} =0$. We here emphasize again that the order-by-order RG invariance holds by virtue of the $\beta$-functions in our scheme. 
As we have done in the one-loop analysis, it is enlightening to discuss the RG invariance in terms of the high-temperature expanded effective potential. Before doing so, we obtain the expression of $\nu^2$ up to $\mathcal{O}(t^2)$. From the $t$-expansion formula (\ref{t-expansion}), it follows that
\begin{align}
\bar{\nu}^2 & \simeq \nu_0^2+\frac{\lambda_0(\nu_0^2-\Sigma)}{16\pi^2}t+\frac{2\lambda_0^2(\nu_0^2-\Sigma)}{(16\pi^2)^2}t^2+\frac{\lambda_0^2(-\nu_0^2+\Sigma)}{(16\pi^2)^2}t +\frac{\lambda_0\Sigma}{16\pi^2}t +2\nu_0^2\gamma_\Phi^{(2)}t. \label{nu2_2L}
\end{align}
Note that $-\lambda_0\Sigma t/16\pi^2$ in the second term is cancelled by $+\lambda_0\Sigma t/16\pi^2$ in the fifth term, which originates from $\beta^{(2)}_{\nu^2}$.
The result would be different if one cancels the whole $\lambda\Sigma/(16\pi^2\nu^2)$ terms in $\beta_{\nu^2}=\beta_{\nu^2}^{(1)}+\beta_{\nu^2}^{(2)}$ from the beginning.
By the cancellation of the $\lambda_0\Sigma t/16\pi^2$ terms, the $\mathcal{O}(1/(16\pi^2))$ term coincides with the corresponding term in the $\overline{\text{MS}}$ scheme. However, the $\mathcal{O}(1/(16\pi^2)^2)$ terms are still different from those in the $\overline{\text{MS}}$ scheme due to the presence of $\Sigma$.
From this demonstration, one could infer that the difference between the two schemes would get smaller at the two-loop level as long as the two-loop corrections are moderate. We will quantify this statement in our numerical analysis. 

As for the quartic coupling $\lambda$ and the scalar field $\varphi$, their running parameters up to $\mathcal{O}(t^2)$ are found to be 
\begin{align}
\bar{\lambda} & \simeq \lambda_0+\frac{3\lambda_0^2}{16\pi^2}t+\frac{9\lambda_0^3}{(16\pi^2)^2}t^2-\frac{6\lambda_0^3}{(16\pi^2)^2}t+4\lambda_0\gamma_\Phi^{(2)}t, \label{lam_2L}\\
\bar{\varphi} &= \exp\left[-\int_{0}^{t}dt'~\gamma_\Phi(t')\right] \varphi_0^{}
\simeq \big(1-\gamma_\Phi^{(2)}t \big)\varphi_0^{}, \label{vphi_2L}
\end{align}
where $\varphi_0^{}=\bar{\varphi}(t=0)$.
The resummed two-loop effective potential in the high-temperature limit is
\begin{align}
\lefteqn{V_{\text{eff}}^{\text{HTE}}(\varphi)=V_0(\varphi)+V_1^{\text{HTE}}(\varphi)+V_2^{\text{HTE}}(\varphi) } \nonumber\\
&\simeq \frac{1}{2}
\Bigg[
-\bigg\{
\nu^2\left(1+\frac{\lambda}{32\pi^2}\ln\frac{T^2}{\bar{\mu}^2}\right) 
-\frac{\lambda^2(-\nu^2+\Sigma)}{2(16\pi^2)^2}\ln^2\frac{T^2}{\bar{\mu}^2}
+\frac{\lambda^2(-\nu^2+\Sigma)}{2(16\pi^2)^2}\ln\frac{T^2}{\bar{\mu}^2}
\bigg\} \nonumber\\
&\hspace{1cm}
+\frac{\lambda T^2}{24}\left(1+\frac{3\lambda}{32\pi^2}\ln\frac{T^2}{\bar{\mu}^2}\right)
+\frac{\lambda^2 T^2}{24(16\pi^2)} 
\left(
2\ln\frac{M^2}{T^2}+1+c_H
\right)
\nonumber\\
&\hspace{1cm}
+\bigg\{
\frac{\lambda(-\nu^2+\Sigma)}{16\pi^2}
+\frac{2\lambda^2(-\nu^2+\Sigma)}{(16\pi^2)^2}\ln\frac{T^2}{\bar{\mu}^2}
\bigg\}c_B
\Bigg]\varphi^2 \nonumber\\
&\quad
-\frac{T}{12\pi}
\bigg[
(M^2)^{3/2}+\frac{3}{4(16\pi^2)}\Big\{\lambda(M^2)^{3/2}+\lambda^2(M^2)^{1/2}\varphi^2\Big\}\ln\frac{T^2}{\bar{\mu}^2}
\bigg] \nonumber\\
&\quad 
+\frac{1}{4!}
\Bigg[
\bigg\{
\lambda+\frac{3\lambda^2}{32\pi^2}\ln\frac{T^2}{\bar{\mu}^2}
+\frac{9\lambda^3}{4(16\pi^2)}\ln^2\frac{T^2}{\bar{\mu}^2}
-\frac{3\lambda^3}{(16\pi^2)^2}\ln\frac{T^2} {\bar{\mu}^2}
\bigg\} 
+\frac{3\lambda^2}{16\pi^2}\left(1+\frac{3\lambda}{16\pi^2}\ln\frac{T^2}{\bar{\mu}^2}\right)c_B
\Bigg]\varphi^4,\label{V2LHTE}
\end{align}
where the terms without explicit $\bar{\mu}$ dependences are only retained up to $\mathcal{O}(1/(16\pi^2))$. 
One can see that the numerical coefficient of $\lambda\ln(T^2/\bar{\mu}^2)/32\pi^2$ in the parenthesis multiplied by the factor $\lambda T^2/24$ in the second line becomes 3 owing to the addition of the two-loop correction, and as a result, this term obeys the one-loop RG equation (\ref{lam_1L})~\cite{Arnold:1992rz}. We also note that all the explicit $\bar{\mu}$ dependences in Eq.~(\ref{V2LHTE}) are absorbed into the running parameters given in Eqs.~(\ref{nu2_2L}), (\ref{lam_2L}), and (\ref{vphi_2L}), resulting in
\begin{align}
V_{\text{eff}}^{\text{HTE}}(\varphi)
&\simeq \frac{1}{2}
\bigg[-\bar{\nu}^2(T)+\frac{\bar{\lambda}(T)T^2}{24}
+\frac{\lambda^2 T^2}{24(16\pi^2)} 
\left(
2\ln\frac{M^2}{T^2}+1+c_H
\right)
 +\frac{\bar{\lambda}(T)(-\bar{\nu}^2(T)+\Sigma)c_B}{16\pi^2}
\bigg]\bar{\varphi}^2 \nonumber\\
&\hspace{1cm}
-\frac{T\big(\bar{M}^2(T)\big)^{3/2}}{12\pi}
+\frac{1}{4!}
\left[
\bar{\lambda}(T)+\frac{3\bar{\lambda}^2(T)c_B}{16\pi^2}
\right]\bar{\varphi}^4,
\end{align}
which is manifestly RG invariant. This $V_{\text{eff}}^{\text{HTE}}$ is common in the $\overline{\text{MS}}$ and our schemes. 
In the $\overline{\text{MS}}$ scheme, however, the explicit $\bar{\mu}$ dependences would remain in the $\mathcal{O}(\lambda^2 \Sigma/(16\pi^2)^2)$ terms, and higher-order terms would be necessary to restore the RG invariance.

\begin{figure}[t]
\center
\includegraphics[width=7.5cm]{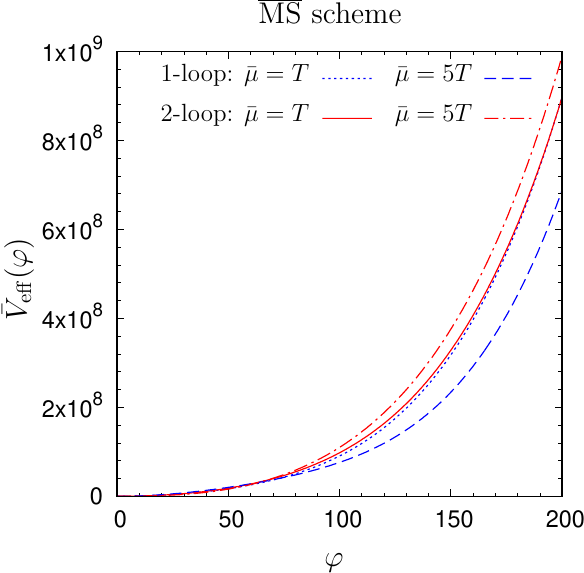}
\hspace{0.5cm}
\includegraphics[width=7.5cm]{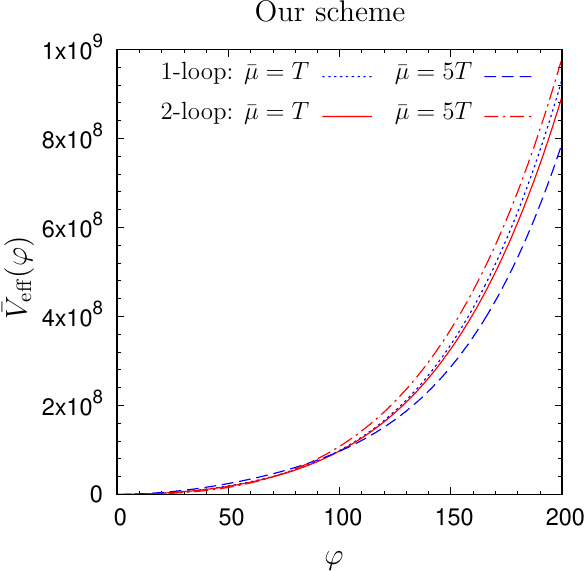}
\caption{Resummed one- and two-loop effective potential with RG improvement in the $\overline{\text{MS}}$ scheme (left) and our scheme (right) at $T=250$. The reference point of the RG running is $\bar{\mu}_0=90$, where we take $v=50$ and $m_\phi=90$ as the inputs, which gives $\lambda\simeq 10$. Note that $\Sigma(T)=\lambda T^2/24$. All the dimensionful parameters are given in units of arbitrary mass dimension.}.
\label{fig:phi4}
\end{figure}

Now, we present numerical results on the $\bar{\mu}$ dependences of the RG-improved effective potentials up to the two-loop level. For practical calculations, we rewrite it as
\begin{align}
\bar{V}_{\text{eff}}(\bar{\varphi};t)=\bar{V}_0(\bar{\varphi};t)+\bar{V}_1(\bar{\varphi};t)+\bar{V}_2(\bar{\varphi};t),\label{RGVeff}
\end{align}
where $t=\ln(\bar{\mu}/\bar{\mu}_0)$ with $\bar{\mu}_0$ representing an initial scale. 
Hereafter, the barred quantities $\bar{\Omega}$, $\bar{\nu}^2$, $\bar{\lambda}$, and $\bar{\varphi}$ are defined as the running parameters which are functions of $t$.
For example, the running parameters obtained by the one-loop $\beta$ functions are, respectively, given by
\begin{align}
\bar{\varphi} &= \varphi\exp\left(-\int_0^t dt'\gamma_\Phi^{(1)}(t')\right)=\varphi,\\
\bar{\lambda} &= \frac{\lambda}{1- \frac{3\lambda}{16\pi^2}t}, \label{barlam1L}\\
\bar{\nu}^2-\Sigma &= \frac{\nu^2-\Sigma}{\left[1-\frac{3\lambda}{16\pi^2}t\right]^{1/3}}, \\
\bar{\Omega} & = \Omega+\frac{(\nu^2-\Sigma)^2}{2\lambda}
\left[
	1-\left(1-\frac{3\lambda}{16\pi^2}t \right)^{1/3}
\right],
\end{align}
where the unbarred parameters are defined at $t=0$. 

In our numerical study, we choose a parameter in which $\Sigma(T)=\lambda(\bar{\mu}_0) T^2/24$ is enhanced to make the difference between $\overline{\text{MS}}$ and our schemes more extensive. One of the examples is shown in Fig.~\ref{fig:phi4}, where the resummed effective potentials with the RG improvement in the $\overline{\text{MS}}$ scheme (left) and our scheme (right) are plotted at $T=250$. The reference point of the RG running is set to $\bar{\mu}_0=90$, and we take $v=50$ and $m_\phi=90$ as the inputs, which corresponds to $\lambda\simeq 10$. All the dimensionful parameters are given in units of arbitrary mass dimension.
`1-loop' denotes $\bar{V}_{\text{eff}}(\bar{\varphi}; t) =  \bar{V}_0(\bar{\varphi}; t)+\bar{V}_1(\bar{\varphi}; t)$ with the one-loop $\beta$-functions in the cases of $\bar{\mu}=T$ (blue, dotted) and $\bar{\mu}=5T$ (blue, dashed), while `2-loop' represents $\bar{V}_{\text{eff}}(\bar{\varphi}; t)=\bar{V}_0(\bar{\varphi}; t)+\bar{V}_1(\bar{\varphi}; t)+\bar{V}_2(\bar{\varphi}; t)$ with the two-loop $\beta$-functions in the cases of $\bar{\mu}=T$ (red, solid) and $\bar{\mu}=5T$ (red, dot-dashed).
One can see that the $\bar{\mu}$ dependence of $\bar{V}_{\text{eff}}$ at the one-loop level in our scheme is generally smaller than that in the $\overline{\text{MS}}$ scheme. This is due to the modified $\beta$-functions in our scheme. 
At the two-loop level, on the other hand, no significant differences between the two schemes are observed, and the $\bar{\mu}$ dependences of $\bar{V}_{\text{eff}}$ are even smaller than the one-loop case in our scheme. As mentioned below Eq.~(\ref{V2LHTE}), the RG invariance in the $\overline{\text{MS}}$ is restored up to $\mathcal{O}(\lambda^2 T^2)$ in the high-temperature limit, which explains our numerical results well.

\begin{figure}[t]
\center
\includegraphics[width=7.5cm]{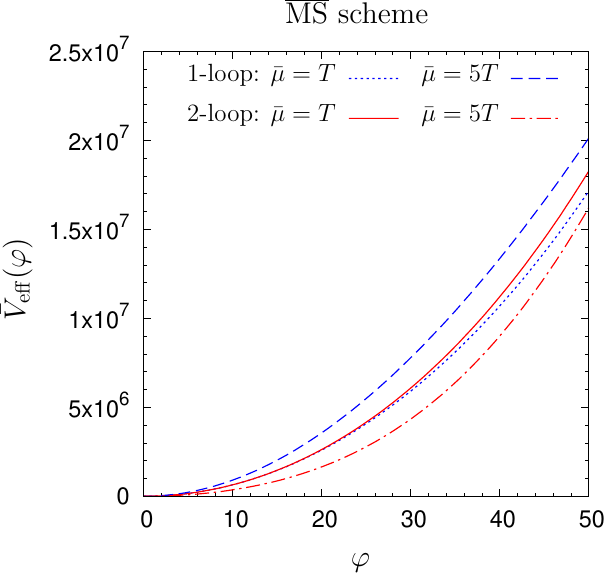}
\hspace{0.5cm}
\includegraphics[width=7.5cm]{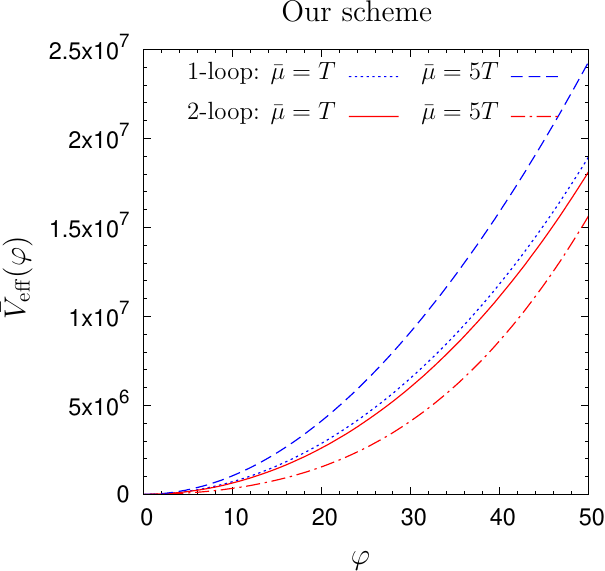}
\caption{The same plots as in Fig.~\ref{fig:phi4} but the range of $\varphi$ is limited to $0 \le \varphi \le 50$. }
\label{fig:phi4_small_phi}
\end{figure}
The $\bar{\mu}$ dependence of the effective potential at the one-loop order obtained by our scheme is smaller than that obtained by the MS-bar
scheme, in the sense that the latter has a larger error in the $\mathcal{D}(V_0+V_1)_{\text{one-loop}}$, as shown in Eqs.~(\ref{DV0_1L})-(\ref{DV1Lms}).
On the other hand, this does not seem to hold for the small-$\varphi$ region in Fig.~\ref{fig:phi4}. To see it easily, we magnify that region and display it in Fig.~\ref{fig:phi4_small_phi}.
This seeming contradiction is caused by an accidental cancellation between the one-loop level error and the original two-loop level one
for the parameters.

In this numerical analysis, $\bar{\mu}=(1-5)T$ is considered to see the $\bar{\mu}$ dependence of $\bar{V}_{\text{eff}}(\varphi; t)$.
The next question is which value of $\bar{\mu}$ is preferable among others. At the one-loop order, for example, it would be useful if there exists $\bar{\mu}$ that can give similar results as the two-loop order. The answer to this question would be very practical when the two-loop effective potential is not at hand. 
For this purpose, we refine the one-loop order $\bar{V}_{\text{eff}}(\varphi; t)$ by judiciously choosing $t$ in the next subsection.

\subsection{Incorporation of higher-order terms}
Following the same spirit of the RG improvement proposed in Refs.~\cite{Kastening:1991gv, Bando:1992np,*Bando:1992wy}, we incorporate higher-order terms utilizing the RG invariance of the effective potential at a given order. 
Here, we focus exclusively on the case of $\bar{V}_{\text{eff}}(\bar{\varphi}; t) =  \bar{V}_0(\bar{\varphi}; t)+\bar{V}_1(\bar{\varphi}; t)$.
When $\bar{V}_{\text{eff}}(\bar{\varphi}, t)$ were exactly $t$ independent, one could choose any $t$ as far as $t$ is below a Landau pole discussed below, and by which it is possible to incorporate a series of dominant higher-order terms via a $\varphi$-dependent $t(\varphi)$.\footnote{When $t$ is $\varphi$ dependent, the running vacuum energy $\bar{\Omega}$ also becomes $\varphi$ dependent so that one cannot simply subtract it from the effective potential. 
} On its trajectory in $t$-$\varphi$ space, $\bar{V}_{\text{eff}}(\bar{\varphi}, t)$ is always flat in the $t$ direction because of the $t$ invariance. As stated above Eq.~(\ref{DV1Lms}), however, the $t$ invariance of $\bar{V}_{\text{eff}}(\bar{\varphi}, t)$ is violated by the two-loop corrections. 
It is thus preferable to choose $t$ such that the truncation error is minimized. With this consideration, we determine $t$ by the condition
\begin{align}
\frac{d \bar{V}_{\text{eff}}(\bar{\varphi};t)}{d t}=\frac{\partial \bar{V}_{\text{eff}}(\bar{\varphi};t)}{\partial t}
& = 0+\frac{1}{2}\frac{\partial\bar{M}^2}{\partial t}\bar{I}(\bar{M})=0,
\label{dVdt_1L}
\end{align}
with
\begin{align}
\frac{\partial \bar{M}^2}{\partial t} = \frac{\bar{\lambda}(\bar{M}^2+\bar{\lambda}\varphi^2)}{16\pi^2},
\end{align}
where the one-loop $\beta$-functions are used. 
From Eq.~(\ref{dVdt_1L}), it follows that
\begin{align}
t(\varphi) &= \frac{8\pi^2}{\bar{M}^2}\bar{I}(\bar{M})_{t=0} 
= \frac{1}{2}
\left[
	\left(\ln\frac{\bar{M}^2}{\bar{\mu}_0^2}-1\right)+\frac{16T^2}{\bar{M}^2}I'_B(\bar{A}^2)
\right].
\label{t-phi}
\end{align}
On the trajectory given by this $t(\varphi)$, $\bar{V}_{\text{eff}}(\bar{\varphi}, t)$ would still be locally flat in the $t$ direction, 
implying that $t(\varphi)$ in Eq.~(\ref{t-phi}) yields the minimal $t$ violation of $\bar{V}_{\text{eff}}(\bar{\varphi}, t)$ among any other choices of $t(\varphi)$. 
In addition to this approximate $t$ invariance, this $t(\varphi)$ copes with two potentially harmful corrections, such as large logarithmic corrections and temperature-dependent power corrections in a general way. 
At zero temperature, Eq.~(\ref{t-phi}) is reduced to $t(\varphi) = \ln(\bar{m}^2/e\bar{\mu}_0^2)/2$ which is connected to the well-known log-resummation scheme $t=\ln(\bar{m}^2/\bar{\mu}_0^2)/2$~\cite{Kastening:1991gv, Bando:1992np,*Bando:1992wy} by changing our initial scale $\mu_0$ to $\mu_0/\sqrt{e}$. At high temperature, on the other hand, Eq.~(\ref{t-phi}) incorporates temperature-dependent power corrections arising from
\begin{align}
\bar{I}^{\text{HTE}}(\bar{M})_{t=0}\simeq \frac{T^2}{12}-\frac{(\bar{M}^2)^{1/2}T}{4\pi}+\frac{\bar{M}^2}{16\pi^2}\ln\frac{\alpha_BT^2}{\bar{\mu}_0^2}+\cdots.
\end{align}
Therefore, $t(\varphi)$ given in Eq.~(\ref{t-phi}) seems to be the best choice for the thermally resummed one-loop effective potential. 
One thing that needs to be noted here is that the truncation error in Eq.~(\ref{dVdt_1L}) is estimated under the assumption that the one-loop $\beta$ functions are used for the running parameters. Instead of this assumption, we could consider the two-loop order running parameters.
In this case, Eq.~(\ref{dVdt_1L}) is modified to
\begin{align}
\frac{d \bar{V}_{\text{eff}}(\bar{\varphi};t)}{d t}& = 0+
\frac{1}{2}\frac{\partial\bar{M}^2}{\partial t}\bar{I}(\bar{M})
-\frac{\bar{\lambda}^2\bar{M}^2\bar{\varphi}^2}{2(16\pi^2)^2}-\frac{\bar{M}^2\Sigma}{16\pi^2}=0,
\label{dVdt_2L}
\end{align}
where 
\begin{align}
\frac{\partial \bar{M}^2}{\partial t}
 &= \frac{\bar{\lambda}(\bar{M}^2+\bar{\lambda}\bar{\varphi}^2-\Sigma)}{16\pi^2}
 -\frac{\bar{\lambda}^2[5(\bar{M}^2+3\bar{\lambda}\bar{\varphi}^2)+\Sigma]}{6(16\pi^2)^2}.
\end{align}
Therefore, the condition of $\bar{I}(\bar{M})=0$ cannot eliminate the whole truncation error when using the two-loop $\beta$ functions.
Although we could, in principle, determine $t(\varphi)$ by the condition (\ref{dVdt_2L}), we still adopt $t(\varphi)$ in Eq.~(\ref{t-phi}) throughout our study due to the benefit described above, i.e., the link to the ordinary log-resummation at zero temperature and $\mathcal{O}(T^2)$ mass resummation at high temperature.

Now we scrutinize if the resummed one-loop effective potential $\bar{V}_{\text{eff}}(\bar{\varphi}; t) =  \bar{V}_0(\bar{\varphi}; t)+\bar{V}_1(\bar{\varphi}; t)$ with the $t$-$\varphi$ relation (\ref{t-phi}) correctly reproduce the fixed-order two-loop effective potential. For this purpose, $\bar{V}_{\text{eff}}(\varphi; t)$ is expanded in powers of $t$,
\begin{align}
\bar{V}_{\text{eff}}(\bar{\varphi}; t) &= \bar{V}_{\text{eff}}(\varphi; 0)
+\frac{\partial \bar{V}_{\text{eff}}(\bar{\varphi}; t)}{\partial t}\bigg|_{t=0}t
+\frac{1}{2}\frac{\partial^2 \bar{V}_{\text{eff}}(\bar{\varphi}; t)}{\partial t^2}\bigg|_{t=0}t^2+\cdots.
\label{appRGVeff}
\end{align}
Let us consider the following two cases
\begin{align}
\bar{V}_{\text{eff}}^{(1)}(\bar{\varphi}; t(\varphi)) &\equiv \bar{V}_0(\bar{\varphi}; t(\varphi))+ \bar{V}_1(\bar{\varphi}; t(\varphi))~\mbox{with the one-loop $\beta$ functions},
\label{barVeff1} \\
\bar{V}_{\text{eff}}^{(2)}(\bar{\varphi}; t(\varphi)) &\equiv \bar{V}_0(\bar{\varphi}; t(\varphi))+ \bar{V}_1(\bar{\varphi}; t(\varphi))~\mbox{with the two-loop $\beta$ functions}.
\label{barVeff2}
\end{align}
Expanding $\bar{V}_{\text{eff}}^{(1)}(\varphi; t(\varphi))$ as the $t$ series, one can find
\begin{align}
\bar{V}_{\text{eff}}^{(1)}(\bar{\varphi}; t(\varphi))
&=\bar{V}_{\text{eff}}^{(1)}(\varphi; 0)
+\frac{\lambda(M^2+\lambda\varphi^2)}{8M^2}\bar{I}^2(M)_{t=0}.
\end{align}
The $\bar{I}^2(M)$ terms are exactly the same as those in $V_2(\varphi)$ shown in Eq.~(\ref{RV2}). 
Similarly, the $t$ series of $\bar{V}_{\text{eff}}^{(2)}(\varphi; t) $ becomes
\begin{align}
\bar{V}_{\text{eff}}^{(2)}(\bar{\varphi}; t(\varphi)) 
&= \bar{V}_{\text{eff}}^{(2)}(\varphi;0)
+\frac{\lambda(M^2+\lambda\varphi^2-\Sigma)}{8M^2}\left(1+\frac{\Sigma}{M^2}\right)\bar{I}^2(M)_{t=0} \nonumber\\
&\quad
-\frac{1}{2}\left(\frac{\lambda^2\varphi^2}{32\pi^2}+\Sigma\right)\bar{I}(M)_{t=0}.
\end{align}
In this case, $\bar{V}_{\text{eff}}^{(2)}(\bar{\varphi};t(\varphi))$ contains not only $\mathcal{O}(\bar{I}^2(M))$ but the $\mathcal{O}(\bar{I}(M))$ terms appearing in $V_2(\varphi)$. One should note that $\Sigma$ terms in $\mathcal{O}(\bar{I}^2(M))$, which are not present in $V_2(\varphi)$, are the consequence of the use of the two-loop $\beta$ functions in $\bar{V}_{\text{eff}}(\bar{\varphi}; t(\varphi))$. From the viewpoint of its RG invariance, such terms can be regarded as higher order terms so that they can be dropped, as we have done in the proof of the RG invariance given in subsection \ref{subsec:RGinv}.  In this sense, $\bar{V}_{\text{eff}}^{(2)}(\bar{\varphi}; t(\varphi))$ correctly resums up to $\mathcal{O}(\bar{I}(M))$. This appears parallel to the leading and next-to-leading logarithmic resummations in the scheme of $t(\varphi)=\ln(\bar{m}^2/\bar{\mu}_0^2)/2$ at zero temperature~\cite{Kastening:1991gv,Bando:1992np,*Bando:1992wy}. 

Before closing this subsection, we discuss the upper limit of $t$. 
As seen from Eq.~(\ref{barlam1L}), $t(\varphi)$ could hit the Landau pole $t_{\text{LP}} = 16\pi^2/3\lambda\simeq 52.6/\lambda$ at which $\bar{\lambda}$ diverges. From the condition $t(\varphi)<t_{\text{LP}}$, it follows that
\begin{align}
\frac{\bar{I}(\bar{M})_{t=0}}{\bar{M}^2}< \frac{2}{3\lambda}.
\label{t_up}
\end{align}
When the $\lambda\times$logarithmic terms are large and/or temperature is significantly high, this condition would not be satisfied. 
Actually, although the parameter set adopted in Fig.~\ref{fig:phi4} illustrates the differences between $\overline{\text{MS}}$ and our schemes clearly, $\lambda\simeq 10$ and $T=250$ are too large to satisfy the condition (\ref{t_up}). In addition to this, since our interest is the case of first-order phase transition required by the gravitational wave generation and EWBG, we extend the $\phi^4$ theory and apply our $t(\varphi)$ to it in the next section. 
%
%
%
\section{$\phi^4$ theory with an additional real scalar}\label{sec:Exphi4}
One of the simplest extensions of the $\phi^4$ theory is to add another real scalar field. 
The bare Lagrangian we consider is defined by
\begin{align}
\mathcal{L}_B &= \sum_{i=1,2}\frac{1}{2}\partial_\mu \Phi_{Bi} \partial^\mu\Phi_{Bi}-V_0(\Phi_{B1},\Phi_{B2}),\\
V_0(\Phi_{B1},\Phi_{B2})&=\Omega_B+\frac{\nu_{B1}^2}{2}\Phi_1^2+\frac{\nu_{B2}^2}{2}\Phi_{B2}^2+\frac{\lambda_{B1}}{4!}\Phi_{B1}^4+\frac{\lambda_{B2}}{4!}\Phi_{B2}^4+\frac{\lambda_{B3}}{4}\Phi_{B1}^2\Phi_{B2}^2,
\end{align}
where two $\mathbb{Z}_2$ symmetries $\Phi_{B1}\to -\Phi_{B1}$ and $\Phi_{B2}\to-\Phi_{B2}$ are imposed to make our analysis simpler.
As we have done in the $\phi^4$ theory, we subtract and add the dominant temperature corrections to the masses of $\Phi_1$ and $\Phi_2$ (denoted as $\Sigma_1$ and $\Sigma_2$) in $\mathcal{L}_R$ and $\mathcal{L}_{\text{CT}}$, respectively. Their explicit forms are given in Appendix~\ref{app:Exphi4}.
For the sake of further simplicity, we also assume that only $\Phi_1$ develops VEV and investigate the thermal phase transition in the $\Phi_1$ direction. 
We define the classical constant background fields and their fluctuation fields as $\Phi_i(x)=\varphi_i+\phi_i(x)$, and VEV of $\phi_1$ is denoted as $v$.

After removing all the divergences of the resummed one-loop effective potential by CTs in Eq.~(\ref{LagCT}) and improving it by RGE (\ref{RGinv}), one would arrive at\footnote{We suppress the $\varphi_2$ dependence by the assumption that only $\Phi_1$ has nonzero VEV.}
\begin{align}
\bar{V}_{\text{eff}}(\bar{\varphi}_1, t) & =\bar{V}_0(\bar{\varphi}_1, t) + \bar{V}_1(\bar{\varphi}_1, t), 
\label{barVeff1L_Exphi4}
\end{align}
where
\begin{align}
\bar{V}_0(\bar{\varphi}_1, t) & = \bar{\Omega}+\frac{1}{2}\left(\bar{\nu}_1^2+\Sigma_1(T)\right)\bar{\varphi}_1^2+\frac{\bar{\lambda}_1}{4!}\bar{\varphi}_1^4, \\
\bar{V}_1(\bar{\varphi}_1, t) & = \sum_{i=1,2}\frac{\bar{M}_i^4}{4(16\pi^2)}\left(\ln\frac{\bar{M}_i^2}{e^{2t}\bar{\mu}_0^2}-\frac{3}{2}\right)+\frac{T^4}{2\pi^2}I_B(\bar{A}_i^2)-\frac{1}{2}\Sigma_1(T)\bar{\varphi}_1^2, 
\end{align}
with  
\begin{align}
\bar{M}_1^2 &= \bar{\nu}_1^2+\Sigma_1(T)+\frac{\bar{\lambda}_1}{2}\bar{\varphi}_1^2,\quad
\bar{M}_2^2  = \bar{\nu}_2^2+\Sigma_2(T)+\frac{\bar{\lambda}_3}{2}\bar{\varphi}_1^2, \\
\Sigma_1(T) & = \frac{T^2}{24}(\lambda_1+\lambda_3), \quad
\Sigma_2(T) = \frac{T^2}{24}(\lambda_2+\lambda_3).
\end{align}
Note that $\Sigma_i(T)$ are given by the parameters at $t=0$ to fulfill the consistency condition as explained in Sec.~\ref{sec:beta},  
Our next step is to refine $\bar{V}_{\text{eff}}(\bar{\varphi}_1,t)$ by incorporating a series of higher-order terms in $\bar{I}(\bar{M}_i)$ via a proper $t$.
As in the $\phi^4$ theory, we choose $t$ for each $\varphi_1$ such that
\begin{align}
\frac{\partial \bar{V}_{\text{eff}}(\bar{\varphi}_1; t)}{\partial t} 
=0+\frac{1}{2}\sum_i\frac{\partial \bar{M}_i^2}{\partial t}\bar{I}(\bar{M}_i)=0,
\end{align}
from which one obtains
\begin{align}
t(\varphi_1) =\frac{8\pi^2 \sum_i\frac{\partial\bar{M}_i^2}{\partial t}\bar{I}(\bar{M}_i)_{t=0} }{\sum_i\bar{M}_i^2\frac{\partial\bar{M}_i^2}{\partial t}}.
\label{t-phi_Exphi4}
\end{align}
Let us approximate $\bar{V}_{\text{eff}}(\bar{\varphi}_1;t)$ in terms of the $t$-expansion and compare it with the two-loop correction to the effective potential~(\ref{RV2_Exphi4}). 
 $\bar{V}_{\text{eff}}^{(1)}(\bar{\varphi}_1, t(\varphi_1))$ defined in Eq.~(\ref{barVeff1}) is found to be
\begin{align}
\bar{V}_{\text{eff}}^{(1)}(\bar{\varphi}_1; t(\varphi_1))
&=\bar{V}_{\text{eff}}^{(1)}(\varphi_1; 0)
+\frac{ \big( \sum_{i=1,2}\alpha_i\bar{I}(M_i)_{t=0} \big)^2 }{8\sum_{i=1,2}\alpha_iM_i^2},
\label{barVeff1_Exphi4}
\end{align}
where $\alpha_i=16\pi^2(\partial \bar{M}_i^2/\partial t)|_{t=0}$, i.e.,
\begin{align}
\alpha_1&=\lambda_1M_1^2+\lambda_3M_2^2+(\lambda_1^2+\lambda_3^2)\varphi_1^2,\\
\alpha_2&=\lambda_3M_1^2+\lambda_2M_2^2+2\lambda_3^2\varphi_1^2.
\end{align}
One can see that $\mathcal{O}(\bar{I}^2(M_i))$ terms in Eq.~(\ref{barVeff1_Exphi4}) do not agree with those in the $V_2(\varphi_1)$ in Eq.~(\ref{RV2_Exphi4}). This is because the single parameter $t$ alone cannot, in principle, incorporate the multiple $\bar{I}^2(M_i)$ terms simultaneously. Only in a special case, such as $|\lambda_1|\gg |\lambda_2|, |\lambda_3|\sim 0$, the $\mathcal{O}(\bar{I}^2(M_i))$ terms in Eq.~(\ref{barVeff1_Exphi4}) would coincide with the corresponding terms of $V_2(\varphi_1)$. 

Similarly, it is straightforward to obtain $\bar{V}_{\text{eff}}^{(2)}(\varphi_1; t(\varphi_1))$ defined in Eq.~(\ref{barVeff2}) as
\begin{align}
\bar{V}_{\text{eff}}^{(2)}(\bar{\varphi}_1; t) 
&= \bar{V}_{\text{eff}}^{(2)}(\varphi_1; 0) \nonumber\\
&\quad 
+\left[\frac{1}{2}\sum_{i=1,2}\frac{\partial \bar{M}_i^2}{\partial t}\bigg|_{t=0}\bar{I}(M_i)_{t=0}
-\sum_{i=1,2}\frac{M_i^2\Sigma_i}{16\pi^2}
-\frac{(\lambda_1^2+\lambda_3^2)M_1^2+2\lambda_3^2M_2^2}{2(16\pi^2)^2}\varphi_1^2
\right]t \nonumber\\
&\quad
+\frac{1}{2}
\left[
	-\sum_{i=1,2}\frac{M_i^2+\Sigma_i}{16\pi^2}\frac{\partial \bar{M}_i^2}{\partial t}\bigg|_{t=0}
\right]
t^2+\cdots
\end{align}
where
\begin{align}
\frac{\partial \bar{M}_1^2}{\partial t}\bigg|_{t=0} &\simeq \frac{\alpha_1}{16\pi^2}-\frac{\lambda_1\Sigma_1+\lambda_3\Sigma_2}{16\pi^2}, \\
\frac{\partial \bar{M}_2^2}{\partial t}\bigg|_{t=0} 
&\simeq \frac{\alpha_2}{16\pi^2}-\frac{\lambda_3\Sigma_1+\lambda_2\Sigma_2}{16\pi^2}.
\end{align}

The $\mathcal{O}(\bar{I}(M_i))$ terms do not agree with those in $V_2(\varphi_1)$ either. Here one may ask wether linear-like terms $(M_i^2)^{1/2}T^3$ in $\bar{V}_{\text{eff}}^{(2)}(\bar{\varphi}_1; t)$ are cancelled or not. As shown below, the answer is positive. Recalling that such terms arise from the high-temperature limit of $\bar{I}^2(M_i)$, i.e, $(T^2/12-(M_i)^{1/2}T/4\pi+\cdots)^2$,  
we take the first derivative of the $\mathcal{O}(\bar{I}^2(M_i))$ terms in $\bar{V}_{\text{eff}}^{(2)}(\bar{\varphi}_1; t)$ with respect to $\bar{I}(M_j)_{t=0}$, which goes like
\begin{align}
\frac{\partial \bar{V}_{\text{eff}}^{(2)}(\bar{\varphi}_1; t)\big|_{\bar{I}^2(M_i)}}{\partial \bar{I}(M_j)_{t=0}}
 & =  \frac{\alpha_j}{4\sum_iM_i^2\alpha_i} 
 \bigg[
 \sum_i \alpha_i \bar{I}(M_i)_{t=0}-2\sum_iM_i^2\Sigma_i
\bigg] \nonumber\\
& \simeq \frac{\alpha_j}{4\sum_iM_i^2\alpha_i} 
\bigg[
\frac{T^2}{12}\sum_i \alpha_i -2\sum_iM_i^2\Sigma_i
\bigg] \nonumber\\
&= \frac{\alpha_jT^2}{48\sum_iM_i^2\alpha_i}\Big[(\lambda_1^2+\lambda_3^2)\varphi_1^2+2\lambda_3^2\varphi_1^2\Big].
\end{align}
Therefore, the linear-like terms are absent in $\bar{V}_{\text{eff}}^{(2)}(\bar{\varphi}_1; t)$.
Although $\mathcal{O}(\bar{I}^2(M_i))$ and $\mathcal{O}(\bar{I}(M_i))$ terms in $\bar{V}_{\text{eff}}^{(1,2)}(\bar{\varphi}_1, t(\varphi_1))$ are different from those in $V_2(\varphi_1)$ in a strict sense, they may still capture the two-loop order corrections that are absent in the resummed one-loop effective potential $V_{\text{eff}}(\bar{\varphi}_1)=V_0(\bar{\varphi}_1)+V_1(\bar{\varphi}_1)$ commonly used in the literature. 
We will quantify to what extent results obtained from $\bar{V}_{\text{eff}}^{(1,2)}(\bar{\varphi}_1, t(\varphi_1))$ are close to those from the resummed two-loop effective potential $\bar{V}_{\text{eff}}(\bar{\varphi}_1; t)=\bar{V}_0(\bar{\varphi}_1; t)+\bar{V}_1(\bar{\varphi}_1; t)+\bar{V}_2(\bar{\varphi}_1; t)$.

\subsection{Numerical anlysis}
Here, we present our numerical results. There are 5 independent parameters in this model, $(\nu_1^2, \nu_2^2, \lambda_1, \lambda_2, \lambda_3)$. 
Some of them can be traded with physical parameters, such as $(v, \nu_2^2, m_{\phi_1}, \lambda_2, m_{\phi_2})$ using the vacuum and mass conditions.
We search for a parameter set that gives the first-order phase transition. 
In particular, we select a case in which differences between the $\overline{\text{MS}}$ and our schemes could be sufficiently large. For that purpose, we take a rather large $\lambda_2$ that enhances $\Sigma_2$.
Moreover, we consider a case in which an imaginary part of the effective potential does not arise near a critical temperature $T_C$, where the effective potential has two degenerate minima. One of parameter sets is given by 
$v=200.0$, $m_{\phi_1}=5.0$, $m_{\phi_2}=125.0$, $\nu_2^2=85.0^2$,  $\lambda_2=5.0$, where those values are given at the initial scale $\bar{\mu}_0$ which is fixed by the condition $t(\varphi_1=v)=0$. It is found that $\bar{\mu}_0\simeq 75.81$ at both the one- and two-loop levels. 
From the input parameters, $\nu_1^2$ and $\lambda_1$ are determined by tadpole and mass conditions at a given order while $\lambda_3$ at the tree level. All the dimensionful parameters are given in units of any mass scale.

\begin{figure}[t]
\begin{center}
\includegraphics[width=7.5cm]{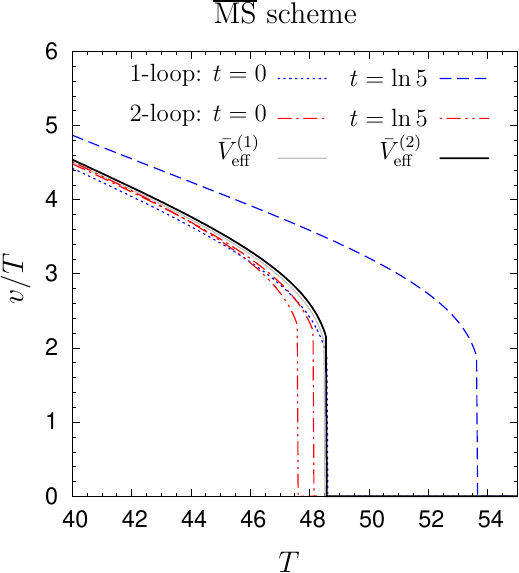}
\hspace{0.5cm}
\includegraphics[width=7.5cm]{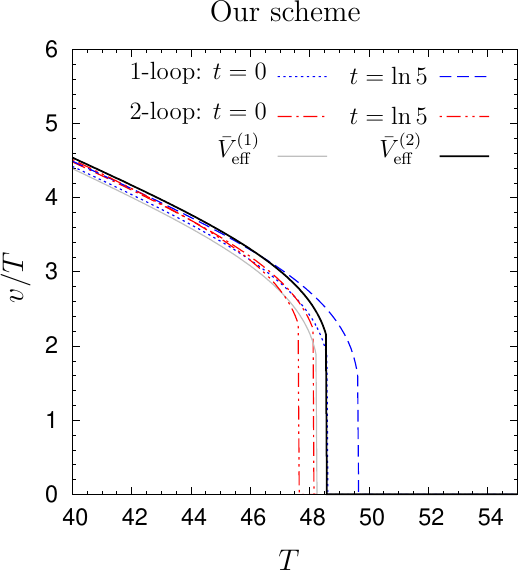}
\caption{$v(T)/T$ as a function of $T$ in the $\overline{\text{MS}}$ scheme (Left) and our scheme (Right). 
}
\label{fig:voverT_T_lam25}
\end{center}
\end{figure}

Fig.~\ref{fig:voverT_T_lam25} shows $v(T)/T$ as functions of the temperature $T$ in the $\overline{\text{MS}}$ (left) and our (right) schemes, respectively. `1-loop' denotes the results using $\bar{V}_{\text{eff}}(\varphi_1;t)=\bar{V}_0(\varphi_1;t)+\bar{V}_1(\varphi_1;t)$ with the one-loop $\beta$-functions in the cases of $t=0$ (blue, dotted) and $t=\ln5$ (blue, dashed), while `2-loop' represents those using $\bar{V}_{\text{eff}}(\varphi_1;t)=\bar{V}_0(\varphi_1;t)+\bar{V}_1(\varphi_1;t)+\bar{V}_2(\varphi_1;t)$ with the two-loop $\beta$-functions in the cases of $t=0$ (red, dot-dashed) and $t=\ln5$ (red, two-dot-dashed). 
The intersections between each curve and horizontal axis correspond to $T_C$.
One can see that $t$ dependence of $T_C$ at the one-loop order in the $\overline{\text{MS}}$ scheme is about 5 times larger than that in our scheme. Such a large $t$ dependence in the $\overline{\text{MS}}$ scheme is reflected by the large RG noninvariance at the order. 
At the two-loop order, on the other hand, the $t$ dependences in both schemes are equally smaller than the one-loop order result in our scheme. 
The significant improvement in the $\overline{\text{MS}}$ scheme is due to the partial restoration of the RG invariance as discussed in the $\phi^4$ theory. As explicitly given in Appendix~\ref{app:Exphi4}, the effective potential follows the RG invariance up to the $\mathcal{O}(\lambda_i^2T^2)$ order in the high-temperature limit. In this parameter choice, the residual RG-violating terms are numerically unimportant, and thus, the $t$ dependence is dominated by the truncation error, leading to similar results in both schemes. 
We also overlay $v(T)/T$ obtained by $\bar{V}_{\text{eff}}^{(1)}(\varphi_1; t(\varphi_1))$ (grey, solid) and  $\bar{V}_{\text{eff}}^{(2)}(\varphi_1; t(\varphi_1))$ (black, thick-solid). It is found that in the two schemes, $v_C/T_C$ in the case of $\bar{V}_{\text{eff}}^{(2)}(\bar{\varphi}_1, t(\varphi_1))$ lie within the two-loop level scale uncertainties, while not in that of $\bar{V}_{\text{eff}}^{(1)}(\bar{\varphi}_1, t(\varphi_1))$.
This demonstration suggests that up to the $\mathcal{O}(\bar{I}(\bar{M}))$ terms are necessary to obtain the results closer to those at the two-loop order. $T_C$ and $v_C/T_C$ in each case are summarized in Table~\ref{tab:summary}.

\begin{figure}[H]
\begin{center}
\includegraphics[width=7.5cm]{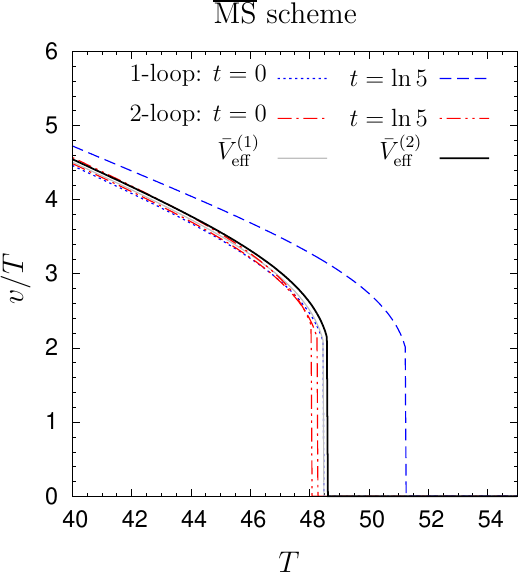}
\hspace{0.5cm}
\includegraphics[width=7.5cm]{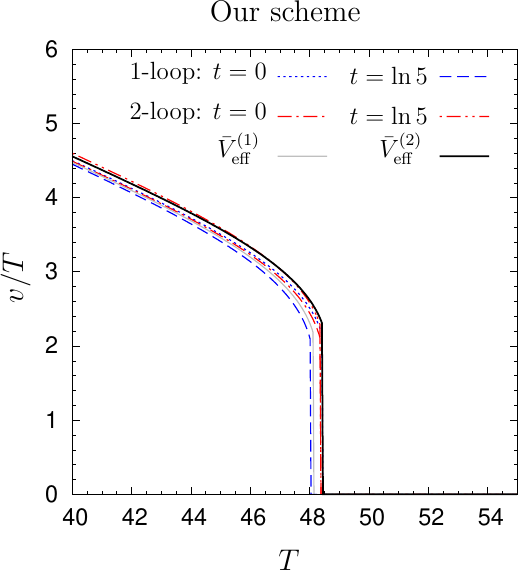} \\[0.5cm]
\includegraphics[width=7.5cm]{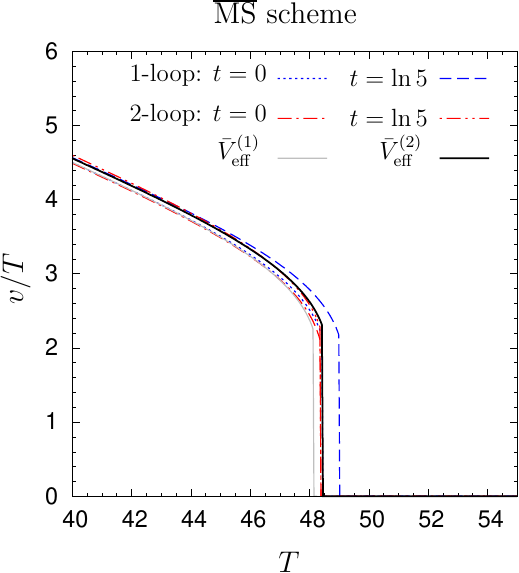}
\hspace{0.5cm}
\includegraphics[width=7.5cm]{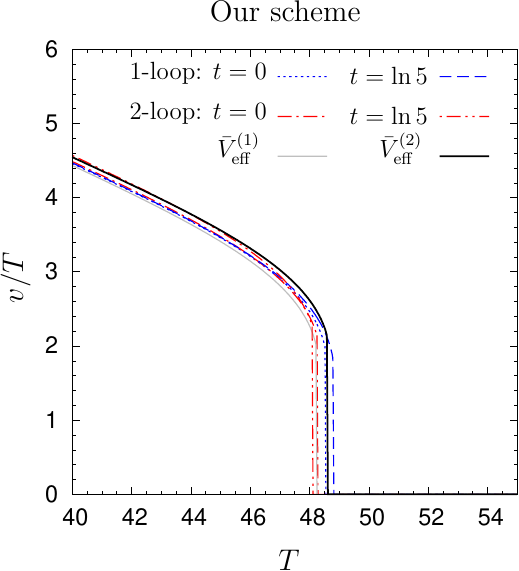} 
\caption{$v(T)/T$ as a function of $T$ in the $\overline{\text{MS}}$ scheme (Left) and our scheme (Right) in the cases of $\lambda_2=3$ (upper plots) and $\lambda_2=1$ (lower plots). The remaining input parameters are the same as in Fig.~\ref{fig:voverT_T_lam25}.
}
\label{fig:voverT_T_lam231}
\end{center}
\end{figure}

As a reference, we also consider the cases of $\lambda_2=1,3$, which give the smaller $\Sigma_2$ compared to the $\lambda_2=5$ case, to see to what extent the two schemes can differ.
In Fig.~\ref{fig:voverT_T_lam231}, $v/T$ is shown as a function of $T$, with the upper plots corresponding to the $\lambda_2=3$ case and the lower ones to the $\lambda_2=1$ case.
The general consequences in those plots are the same as in Fig.~\ref{fig:voverT_T_lam25}, but the differences between the two schemes in the one-loop order results get smaller as $\lambda_2$ becomes smaller. $T_C$ and $v_C/T_C$ in all the cases are listed in Table~\ref{tab:summary}.

\begin{table}[t]
\center
\begin{tabular}{|c|c|c|c|c|c|c|c|c|c|}
\hline
 & \multicolumn{4}{c|}{$\overline{\text{MS}}$ scheme} & \multicolumn{4}{c|}{Our scheme} \\ \hline
 & \multicolumn{8}{c|}{$T_C$}  \\ \hline
 & 1-loop & 2-loop & $\bar{V}_{\text{eff}}^{(1)}$ & $\bar{V}_{\text{eff}}^{(2)}$ & 1-loop & 2-loop & $\bar{V}_{\text{eff}}^{(1)}$ & $\bar{V}_{\text{eff}}^{(2)}$ \\ \hline
$\lambda_2=5$ & $48.6-53.6$ & $47.6-48.1$ & $48.5$ & $48.6$ & $48.6-49.6$ & $47.6-48.1$ & $48.2$ & $48.6$ \\ 
$\lambda_2=3$ & $48.5-51.2$ & $48.1-48.3$ & $48.5$ & $48.6$ & $48.5-48.8$ & $48.1-48.3$ & $48.2$ & $48.6$ \\ 
$\lambda_2=1$ & $48.4-49.0$ & $48.4$ & $48.1$ & $48.4$ & $48.0-48.4$ & $48.4$ & $48.1$ & $48.4$ \\ 
\hline \hline
 & \multicolumn{8}{c|}{$v_C/T_C$}  \\ \hline
 & 1-loop & 2-loop & $\bar{V}_{\text{eff}}^{(1)}$ & $\bar{V}_{\text{eff}}^{(2)}$ & 1-loop & 2-loop & $\bar{V}_{\text{eff}}^{(1)}$ & $\bar{V}_{\text{eff}}^{(2)}$ \\ \hline
$\lambda_2=5$ & $1.9$ & $2.1-2.3$ & $2.1$ & $2.2$ & $1.6-1.9$ & $2.2-2.3$ & $1.9$ & $2.2$ \\ 
$\lambda_2=3$ & $2.0-2.1$ & $2.1-2.3$ & $2.1$ & $2.1$ & $1.8-2.0$ & $2.1-2.3$ & $2.0$ & $2.1$ \\ 
$\lambda_2=1$ & $2.2$ & $2.1-2.3$ & $2.3$ & $2.3$ & $2.1-2.2$ & $2.1-2.3$ & $2.2$ & $2.3$ \\ 
\hline
\end{tabular}
\caption{
The values of $v_C$ and $T_C$ are summarized in the case of $\lambda_2=1,3,5$ for the $\overline{\text{MS}}$ and our renormalization schemes. Here, `1-loop' and `2-loop' denote the values obtained by the $t$-dependent effective potential at the one- and two-loop orders, respectively, for $t$ in the range of $0<t<\ln 5$.
}.
\label{tab:summary}
\end{table}

\section{Conclusion and discussions}\label{sec:conclusion}
We have presented our RG improvement for the thermally resummed effective potentials in detail. In our method, $\beta$-functions are defined in the resummed theory, and thus, the order-by-order RG invariance of the effective potential holds consistently, which is in stark contrast to the case of $\overline{\text{MS}}$ scheme. As a simple example, we applied our method to the $\phi^4$ theory and made a comparison with the $\overline{\text{MS}}$ scheme both analytically and numerically. At the one-loop order, our scheme generally gives less scale dependences than the $\overline{\text{MS}}$ scheme does. At the two-loop order, however, the differences between the two schemes are not pronounced since the scale invariance is restored up to $\mathcal{O}(\lambda^2T^2)$ in the $\overline{\text{MS}}$ scheme.
Our numerical study also exemplifies the case that the scale dependence in the $\overline{\text{MS}}$ scheme can become smaller than that in our scheme due to the accidental cancellation between the RG-noninvariant terms and truncation errors. 
This demonstration illustrates the need to exercise caution when interpreting the scale dependence.
We also proposed the refinement for the resummed one-loop effective potential in which the one-loop function ($\bar{I}(\bar{M})$) as a whole is resummed by fully utilizing the RG invariance. Because of its general form, the potentially dangerous large logarithmic terms and power corrections of temperature are simultaneously tamed.
Moreover, this method is less sensitive to truncation errors than any other choice.  

We also discussed the first-order phase transition in the $\phi^4$ theory augmented by another real scalar field. 
We showed that the scale dependence of $T_C$ obtained by the resummed one-loop effective potential is much smaller than that in the $\overline{\text{MS}}$ scheme owing to the modified $\beta$-functions. At the two-loop order, however, both schemes are equally good as in the $\phi^4$ theory. Our numerical study shows that the resummed one-loop effective potential with the two-loop $\beta$-functions ($\bar{V}_{\text{eff}}^{(2)}$) can yield the same $v_C/T_C$ as those in the two-loop order calculations within their scale uncertainties, implying that the dominant two-loop order contributions are incorporated into $\bar{V}_{\text{eff}}^{(2)}$ to a good approximation. 
This suggests that $\bar{V}_{\text{eff}}^{(2)}$ could be practically useful when the full two-loop effective potentials are not at hand. 

In Ref.~\cite{Funakubo:2012qc}, we show the renormalizability of resummed two-loop effective potentials without resorting to HTE in abelian gauge theories. It would be interesting to clarify whether our method also leads to the same conclusion obtained here. We leave this to future research~\cite{FS}.


\appendix

\section{Counterterms and $\beta$-functions in the resummed theories}
%
%
\subsection{$\phi^4$ theory}\label{app:phi4}
We divide the bare Lagrangian (\ref{bLag_phi4}) into the renormalized part and counterterms:
\begin{align}
\mathcal{L}_B = \mathcal{L}_R +\mathcal{L}_{\text{CT}},
\end{align}
where
\begin{align}
\mathcal{L}_R & = \frac{1}{2}\partial_\mu \Phi \partial^\mu\Phi-\Omega+\frac{\nu^2}{2}\Phi^2-\frac{\lambda\mu^\epsilon}{4!}\Phi^4, \\
\mathcal{L}_{\text{CT}} & = \frac{1}{2}(Z_\Phi-1)\partial_\mu \Phi \partial^\mu \Phi-\delta\Omega+\frac{\delta\nu^2}{2}\Phi^2-\frac{\delta \lambda \mu^\epsilon}{4!}\Phi^4.\label{CTphi4}
\end{align}
The relationships between the bare and renormalized parameters are, respectively, given by
\begin{align}
\Phi_B = Z_\Phi^{1/2}\Phi,\quad \nu_B^2 = Z_\Phi^{-1}(\nu^2+\delta\nu^2),\quad 
\lambda_B\mu^{-\epsilon}  = Z_\Phi^{-2}(\lambda+\delta\lambda),\quad \Omega_B\mu^\epsilon = \Omega+\delta\Omega.\label{reno_consts_phi4}
\end{align}
$\mathcal{L}_R$ and $\mathcal{L}_{\text{CT}}$ in the resummed $\phi^4$ theory are modified as
\begin{align}
\mathcal{L}_R
& = \frac{1}{2}\partial_\mu \Phi \partial^\mu \Phi-\Omega+\frac{\nu^2-\Sigma(T)}{2}\Phi^2-\frac{\lambda\mu^\epsilon}{4!}\Phi^4, \\
\mathcal{L}_{\text{CT}} & = \frac{1}{2}(Z_\Phi-1)\partial_\mu \Phi \partial^\mu\Phi -\delta\Omega+\frac{\delta\nu^2+\Sigma(T)}{2}\Phi^2-\frac{\delta \lambda \mu^\epsilon}{4!}\Phi^4.
\end{align}
Note that the relations in Eq.~(\ref{reno_consts_phi4}) remain intact. 
When the spontaneous symmetry breaking occurs, the scalar field is shifted as $\Phi(x)=\varphi+\phi(x)$.
As in the ordinary perturbation theory, CTs are perturbatively expanded as
\begin{align}
\delta \Omega &= \delta^{(1)} \Omega +  \delta^{(2)} \Omega+\cdots, \label{CTOmega} \\
\delta \nu^2 &= \delta^{(1)} \nu^2 +  \delta^{(2)} \nu^2+\cdots, \label{CTnu2} \\
\delta \lambda &= \delta^{(1)} \lambda +  \delta^{(2)} \lambda +\cdots,\label{CTlam} \\
Z_\Phi & = 1+z_\Phi^{(1)}+z_\Phi^{(2)} +\cdots, \label{CTZPhi}
\end{align}
and determined order-by-order in the resummed perturbation theory.
At the one-loop level, CTs are given in Eq.~(\ref{1LCT}) and at the two-loop level, one can find
\begin{align}
\delta^{(2)}\Omega & = \frac{\lambda(\nu^2-\Sigma)^2}{2(16\pi^2)^2}\frac{1}{\epsilon^2}+\frac{(\nu^2-\Sigma)\Sigma}{16\pi^2}\frac{1}{\epsilon}, \\
\delta^{(2)}\nu^2 &= \frac{\lambda^2(\nu^2-\Sigma)}{(16\pi^2)^2}\left(\frac{2}{\epsilon^2}-\frac{1}{2\epsilon}\right)
	+\frac{\lambda\Sigma}{16\pi^2}\frac{1}{\epsilon}, \\
\delta^{(2)}\lambda &= \frac{3\lambda^2}{(16\pi^2)^2}\left(\frac{3}{\epsilon^2}-\frac{1}{\epsilon} \right), \\
z_\Phi^{(2)} &= -\frac{\lambda^2}{12(16\pi^2)^2}\frac{1}{\epsilon}.
\end{align}
It would be instructive to show the derivation of $\beta_{\nu^2}^{(1)}$ and $\beta_\Omega^{(1)}$ in more detail. The bare mass $\nu_B^2$ is expressed as
\begin{align}
\nu_B^2 = \nu^2\left(1+\sum_{n=1}^\infty\frac{b_n(\lambda)}{\epsilon^n} \right)+\Sigma(T)\sum_{n=1}^\infty\frac{\tilde{b}_n(\lambda)}{\epsilon^n}.
\end{align}
Applying $d/dt=\mu d/d\mu$ in both sides, one gets
\begin{align}
0 =\nu^2\beta_{\nu^2}^{(\epsilon)}\left(1+\sum_{n=1}^\infty\frac{b_n(\lambda)}{\epsilon^n}\right)
+(\nu^2+\Sigma(T))\sum_{n=1}^\infty\frac{\beta_\lambda^{(\epsilon)}}{\epsilon^n}\frac{db_n(\lambda)}{d\lambda}
+\frac{d\Sigma(T)}{d\lambda} \beta_\lambda^{(\epsilon)} \sum_{n=1}^\infty \frac{\tilde{b}_n(\lambda)}{\epsilon^n},
\end{align}
where $\beta_\lambda^{(\epsilon)}=d\lambda/dt=\sum_{n=0}^\infty x_n\epsilon^n$. Since $x_n=0$ for $n\ge 2$, $\beta_\lambda^{(\epsilon)}=x_0+x_1\epsilon=x_0-\lambda \epsilon$, which leads to
\begin{align}
\nu^2\beta_{\nu^2} = \lambda\nu^2\frac{db_1(\lambda)}{d\lambda} 
+\lambda\Sigma(T)\frac{d\tilde{b}_1(\lambda)}{d\lambda}
+ \lambda\tilde{b}_1\frac{d\Sigma(T)}{d\lambda}.
\end{align}
At the one-loop level, one obtains $b_1(\lambda)=-\tilde{b}_1(\lambda)=\lambda/16\pi^2$ from Eq.~(\ref{V1phi4}). One finally arrives at
\begin{align}
\nu^2\beta_{\nu^2}^{(1)} = \frac{\lambda(\nu^2-\Sigma(T))}{16\pi^2}-\frac{\lambda^2}{16\pi^2}\frac{d\Sigma(T)}{d\lambda}.
\end{align}
If we adopt the resummation method in which $d\Sigma(T)/dt\neq0$, the last term should be kept. 
However, such a term would not preserve the RG invariance at the one-loop order.
In our resummation method with the consistency condition, on the other hand,  $\beta_{\nu^2}^{(1)}$ is reduced to
\begin{align}
\nu^2\beta_{\nu^2}^{(1)} = \frac{\lambda(\nu^2-\Sigma(T))}{16\pi^2}.
\end{align}

Now we move on to derive $\beta_\Omega^{(1)}$. The bare vacuum energy is expressed as
\begin{align}
\Omega_B\mu^\epsilon = \Omega+\sum_{n=1}^\infty\frac{\omega_n(\lambda)}{\epsilon^n}.
\end{align}
where the $\lambda$ dependence of $\omega_n(\lambda)$ arise from $\Sigma(T)$.
Taking the $t$-derivative of both sides, one finds
\begin{align}
\epsilon\Omega_B\mu^\epsilon=
\epsilon\left[\Omega+\sum_{n=1}^\infty\frac{\omega_n(\lambda)}{\epsilon^n} \right] = 
\beta_\Omega^{(\epsilon)}+\sum_{n=1}^\infty \frac{1}{\epsilon^n}\mu\frac{d\omega_n(\lambda)}{d\mu},
\end{align}
where $\beta_\Omega^{(\epsilon)}=\mu d\Omega/d\mu$. With $\beta_\Omega^{(\epsilon)}=\sum_{n=0}^\infty d_n\epsilon^n$ and $\beta_\lambda^{(\epsilon)}=x_0-\lambda \epsilon$ and taking $\epsilon\to 0$, 
\begin{align}
\beta_\Omega = \lim_{\epsilon\to0}\beta_\Omega^{(\epsilon)} = d_0 = \omega_1+\lambda \frac{d\omega_1(\lambda)}{d\lambda},
\end{align}
where the second term is induced by the running of $\Sigma(T)$. Thus, such a term should be discarded if the consistency condition applies, and we are left with
\begin{align}
\beta_\Omega^{(1)}  = \frac{(\nu^2-\Sigma)^2}{32\pi^2}.
\end{align}

%
%
\subsection{$\phi^4$ theory with additional scalar}\label{app:Exphi4}
Following the same procedure in the $\phi^4$ theory, the renormalized Lagrangian and CTs after the thermal resummation are, respectively, given by 
\begin{align}
\mathcal{L}_R & = 
\sum_{i=1,2}\frac{1}{2}\partial_\mu \Phi_{i} \partial^\mu\Phi_{i}-V_0(\Phi_{1},\Phi_{2}),\\
\mathcal{L}_{\text{CT}} & = \frac{1}{2}\sum_i(Z_{\Phi_i}-1)\partial_\mu \Phi_i\partial^\mu \Phi_i-\delta V_0(\Phi_{1},\Phi_{2}),
\end{align}
where
\begin{align}
V_0(\Phi_{1},\Phi_{2})&=\Omega+\frac{\nu_{1}^2+\Sigma_1(T)}{2}\Phi_1^2+\frac{\nu_{2}^2+\Sigma_2(T)}{2}\Phi_{2}^2+\frac{\lambda_{1}}{4!}\Phi_{1}^4+\frac{\lambda_{2}}{4!}\Phi_{2}^4+\frac{\lambda_{3}}{4}\Phi_{1}^2\Phi_{2}^2, \\
\delta V_0(\Phi_{1},\Phi_{2})&=\delta\Omega\mu^{-\epsilon}+\frac{\delta\nu_{1}^2-\Sigma_1(T)}{2}\Phi_1^2+\frac{\delta\nu_{2}^2-\Sigma_2(T)}{2}\Phi_{2}^2+\frac{\delta\lambda_{1}\mu^{\epsilon}}{4!}\Phi_{1}^4+\frac{\delta\lambda_{2}\mu^{\epsilon}}{4!}\Phi_{2}^4+\frac{\delta\lambda_{3}\mu^\epsilon}{4}\Phi_{1}^2\Phi_{2}^2.
\label{LagCT}
\end{align}
The relationships between the bare and renormalized parameters are 
\begin{align}
\Phi_{iB} &= Z_{\Phi_i}^{1/2}\Phi_i,\quad \nu_i^2 = Z_{\Phi_i}^{-1}(\nu_i^2+\delta\nu_i^2),\quad 
\lambda_{iB}\mu^{-\epsilon} = Z_{\Phi_i}^{-2}(\lambda_i+\delta\lambda_i),\quad i=1,2 \\
\lambda_{3B}\mu^{-\epsilon} & = Z_{\Phi_1}^{-1}Z_{\Phi_2}^{-1}(\lambda_3+\delta\lambda_3),\quad
\Omega_B\mu^\epsilon = \Omega+\delta\Omega.
\end{align}
As in Eqs.~(\ref{CTOmega})-(\ref{CTZPhi}), CTs are determined order by order in the resummed perturbation theory. The one-loop order CTs are, respectively, given by
\begin{align}
\delta^{(1)}\Omega &= \frac{(\nu_1^2+\Sigma_1)^2+(\nu_2^2+\Sigma_2)^2}{2(16\pi^2)}\frac{1}{\epsilon}, \\
\delta^{(1)}\nu_1^2 &= \frac{\lambda_1(\nu_1^2+\Sigma_1)+\lambda_3(\nu_2^2+\Sigma_2) }{16\pi^2}\frac{1}{\epsilon}, \\
\delta^{(1)}\lambda_1 &= \frac{3(\lambda_1^2+\lambda_3^2)}{16\pi^2}\frac{1}{\epsilon}, \\
z_{\Phi_1}^{(1)} & = 0.
\end{align}
while the two-loop order CTs are
\begin{align}
\delta^{(2)}\Omega &= \frac{\lambda_1(\nu_1^2+\Sigma_1)^2+\lambda_2(\nu_2^2+\Sigma_2)^2+2\lambda_3(\nu_1^2+\Sigma_1)(\nu_2^2+\Sigma_2)}{2(16\pi^2)^2}\frac{1}{\epsilon^2} \nonumber\\
&\quad 
	-\frac{\Sigma_1(\nu_1^2+\Sigma_1)+\Sigma_2(\nu_2^2+\Sigma_2)}{16\pi^2}\frac{1}{\epsilon}, \\
\delta^{(2)}\nu_1^2 &= \frac{2(\lambda_1^2+\lambda_3^2)(\nu_1^2+\Sigma_1)+\lambda_3(\lambda_1+\lambda_2+2\lambda_3)(\nu_2^2+\Sigma_2)}{(16\pi^2)^2}
\frac{1}{\epsilon^2} \nonumber\\
&\quad 
-\left[
\frac{(\lambda_1^2+\lambda_3^2)(\nu_1^2+\Sigma_1)+2\lambda_3^2(\nu_2^2+\Sigma_2)}{2(16\pi^2)^2}
+\frac{\lambda_1\Sigma_1+\lambda_3\Sigma_2}{16\pi^2}
\right]\frac{1}{\epsilon}, \\
\delta^{(2)}\lambda_1 &= \frac{3(3\lambda_1^3+4\lambda_1\lambda_3^2+\lambda_2\lambda_3^2+4\lambda_3^3)}{(16\pi^2)^2}\frac{1}{\epsilon^2} 	-\frac{3[\lambda_1(\lambda_1^2+\lambda_3^2)+2\lambda_3^3]}{(16\pi^2)^2}\frac{1}{\epsilon}, \\
z_{\Phi_1}^{(2)} & = -\frac{\lambda_1^2+3\lambda_3^2}{12(16\pi^2)^2}\frac{1}{\epsilon}.
\end{align}

The classical constant background fields and their fluctuation fields are denoted as $\Phi_i(x)=\varphi_i+\phi_i(x)$.
After the renormalization in our scheme, the resummed effective potential up to the two-loop level is 
\begin{align}
V_0(\varphi_1) & = \Omega+\frac{1}{2}\left(\nu_1^2+\Sigma_1(T)\right)\varphi_1^2+\frac{\lambda_1}{4!}\varphi_1^4, \\
V_1(\varphi_1) & = \sum_{i=1,2}\frac{M_i^4}{4(16\pi^2)}\left(\ln\frac{M_i^2}{\bar{\mu}^2}-\frac{3}{2}\right)+\frac{T^4}{2\pi^2}I_B(A_i^2)-\frac{1}{2}\Sigma_1(T)\varphi_1^2, \\
V_2(\varphi_1) & = -\frac{\varphi_1^2}{4}\left[\frac{\lambda_1^2}{3}\tilde{H}(M_1)+\lambda_3^2\tilde{H}(M_1, M_2, M_2)\right] \nonumber\\
&\quad +\frac{1}{8}\Big[\lambda_1\bar{I}^2(M_1)+\lambda_2\bar{I}^2(M_2)+2\lambda_3\bar{I}(M_1)\bar{I}(M_2) \Big]
-\frac{1}{2}\Big[\Sigma_1\bar{I}(M_1)+\Sigma_2\bar{I}(M_2)\Big],
\label{RV2_Exphi4} 
\end{align}
where $\tilde{H}(M_1)=\tilde{H}(M_1,M_1,M_1)$ defined in Eq.~(\ref{tilH123}), $A_i=M_i/T$ and 
\begin{align}
M_1^2 &= \nu_1^2+\Sigma_1(T)+\frac{\lambda_1}{2}\varphi_1^2,\quad
M_2^2  = \nu_2^2+\Sigma_2(T)+\frac{\lambda_3}{2}\varphi_1^2, \\
\Sigma_1(T) & = \frac{T^2}{24}(\lambda_1+\lambda_3), \quad
\Sigma_2(T) = \frac{T^2}{24}(\lambda_2+\lambda_3).
\end{align}
As is the $\phi^4$ theory case, one can verify the order-by-order RG invariance of the above effective potential in terms of the $\beta$-functions in our scheme. One-loop $\beta$-functions are given by
\begin{align}
\gamma_{\Phi_1}^{(1)} & =\gamma_{\Phi_2}^{(1)} = 0, \\
\beta_{\Omega}^{(1)} & = \frac{1}{32\pi^2}\Big[(\nu_1^2+\Sigma_1)^2+(\nu_2^2+\Sigma_2)^2\Big], \\
\nu_1^2\beta_{\nu_1^2}^{(1)} & = \frac{1}{16\pi^2}
\Big[
	\lambda_1(\nu_1^2+\Sigma_1)+\lambda_3(\nu_2^2+\Sigma_2)
\Big], \\
\nu_2^2\beta_{\nu_2^2}^{(1)} 
& = \frac{1}{16\pi^2}\Big[\lambda_3(\nu_1^2+\Sigma_1)+\lambda_2(\nu_2^2+\Sigma_2)\Big], \\
\beta_{\lambda_1}^{(1)} & = \frac{3}{16\pi^2} (\lambda_1^2+\lambda_3^2), \\
\beta_{\lambda_2}^{(1)} & = \frac{3}{16\pi^2} (\lambda_2^2+\lambda_3^2), \\
\beta_{\lambda_3}^{(1)} 
& = \frac{\lambda_3(\lambda_1+\lambda_2+4\lambda_3)}{16\pi^2},
\end{align}
and two-loop $\beta$-functions we need are
\begin{align}
\beta_{\Omega}^{(2)} 
& = -\frac{1}{16\pi^2}\Big[(\nu_1^2+\Sigma_1)\Sigma_1+(\nu_2^2+\Sigma_2)\Sigma_2 \Big], \\
\nu_1^2\beta_{\nu_1^2}^{(2)} 
& = -\frac{1}{(16\pi^2)^2}
\Big[(\lambda_1^2+\lambda_3^2)(\nu_1^2+\Sigma_1)+2\lambda_3^2(\nu_2^2+\Sigma_2)\Big]
-\frac{\lambda_1\Sigma_1+\lambda_3\Sigma_2}{16\pi^2}
+2\nu_1^2\gamma_{\Phi_1}^{(2)},\\
\nu_2^2\beta_{\nu_2^2}^{(2)} 
& = -\frac{1}{(16\pi^2)^2}
\Big[(\lambda_2^2+\lambda_3^2)(\nu_2^2+\Sigma_2)+2\lambda_3^2(\nu_1^2+\Sigma_1)\Big]
-\frac{\lambda_2\Sigma_2+\lambda_3\Sigma_2}{16\pi^2}
+2\nu_2^2\gamma_{\Phi_2}^{(2)},\\
\beta_{\lambda_1}^{(2)} 
& = \frac{-6}{(16\pi^2)^2}\Big[\lambda_1(\lambda_1^2+\lambda_3^2)+2\lambda_3^3\Big]
+4\lambda_1\gamma_{\Phi_1}^{(2)}, \\
\beta_{\lambda_2}^{(2)} 
& = \frac{-6}{(16\pi^2)^2}\Big[\lambda_2(\lambda_2^2+\lambda_3^2)+2\lambda_3^3\Big]
+4\lambda_2\gamma_{\Phi_2}^{(2)}, \\
\gamma_{\Phi_1}^{(2)} & = \frac{\lambda_1^2+3\lambda_3^2}{12(16\pi^2)^2}, \\
\gamma_{\Phi_2}^{(2)} & = \frac{\lambda_2^2+3\lambda_3^2}{12(16\pi^2)^2}.
\end{align}
As in the previous special case, one can find
\begin{align}
\mathcal{D}V_0|_{\text{one-loop}} & = \beta_{\Omega}^{(1)}+\frac{\nu_1^2}{2}\beta_{\nu_1^2}^{(1)}\varphi_1^2+\frac{1}{4!}\beta_{\lambda_1}^{(1)}\varphi_1^4 = \frac{M_1^4+M_2^4}{2(16\pi^2)},\\
\mathcal{D}V_1|_{\text{one-loop}} &=\mu\frac{\partial V_1}{\partial \mu}=-\frac{M_1^4+M_2^4}{2(16\pi^2)},
\end{align}
which verifies that $\mathcal{D}(V_0+V_1)|_{\text{one-loop}}=0$.

Now we consider $V_{\text{eff}}^{\text{HTE}}(\varphi_1)$.
\begin{align}
V_{\text{eff}}^{\text{HTE}}(\varphi_1) & = V_0(\varphi_1)+V_1^{\text{HTE}}(\varphi_1) \nonumber\\
&\simeq
\frac{1}{2}
\bigg[
\left\{\nu_1^2+\frac{\lambda_1(\nu_1^2+\Sigma_1)+\lambda_3(\nu_2^2+\Sigma_2)}{32\pi^2}\ln\frac{T^2}{\bar{\mu}^2} \right\}
+\frac{(\lambda_1+\lambda_3)T^2}{24} \nonumber\\
&\hspace{1cm}
	+\frac{1}{16\pi^2}\Big\{\lambda_1(\nu_1^2+\Sigma_1)+\lambda_3(\nu_2^2+\Sigma_2)\Big\}c_B
\bigg]\varphi_1^2-\frac{T\big( (M_1^2)^{3/2}+(M_2^2)^{3/2} \big) }{12\pi} \nonumber\\
&\quad+\frac{1}{4!}
\left[
\left(\lambda_1+\frac{3(\lambda_1^2+\lambda_3^2)}{32\pi^2}\ln\frac{T^2}{\bar{\mu}^2}\right)
+\frac{3(\lambda_1^2+\lambda_3^2)c_B}{16\pi^2}
\right]\varphi_1^4+\cdots, \nonumber\\
&=\frac{1}{2}
\bigg[
\bar{\nu}_1^2(T)+\frac{(\lambda_1+\lambda_3)T^2}{24}
+\frac{1}{16\pi^2}\Big\{\lambda_1(\nu_1^2+\Sigma_1)+\lambda_3(\nu_2^2+\Sigma_2)\Big\}c_B
\bigg]\varphi_1^2 \nonumber\\
&\quad 
-\frac{T\big( (M_1^2)^{3/2}+(M_2^2)^{3/2} \big) }{12\pi}
+\frac{1}{4!}
\left[\bar{\lambda}_1(T)+\frac{3(\lambda_1^2+\lambda_3^2)c_B}{16\pi^2}
\right]\varphi_1^4+\cdots,
\end{align}
where $\bar{\nu}_1^2$ and $\bar{\lambda}_1$ are the running parameters in our scheme. To see difference between the $\overline{\text{MS}}$ and our schemes, we rewrite $V_{\text{eff}}^{\text{HTE}}(\varphi_1)$ by taking $\Sigma_1=(\lambda_1+\lambda_3)T^2/24$ and $\Sigma_2=(\lambda_2+\lambda_3)T^2/24$, resulting in 
\begin{align}
V_{\text{eff}}^{\text{HTE}}(\varphi_1)
&= \frac{1}{2}
\bigg[
\bar{\nu}_1^2(T)|_{\overline{\text{MS}}}
+\frac{T^2}{24}\left\{ \lambda_1\left(1+\frac{\lambda_1+\lambda_3}{32\pi^2}\ln\frac{T^2}{\bar{\mu}^2}\right)
+\lambda_3\left(1+\frac{\lambda_2+\lambda_3}{32\pi^2}\ln\frac{T^2}{\bar{\mu}^2}\right) \right\} \nonumber\\
&\hspace{1cm}
+\frac{1}{16\pi^2}\Big\{\lambda_1(\nu_1^2+\Sigma_1)+\lambda_3(\nu_2^2+\Sigma_2)\Big\}c_B
\bigg]\varphi_1^2 \nonumber\\
&\quad
-\frac{T\big( (M_1^2)^{3/2}+(M_2^2)^{3/2} \big) }{12\pi}
+\frac{1}{4!}
\left[
\bar{\lambda}_1(T)+\frac{3(\lambda_1^2+\lambda_3^2)c_B}{16\pi^2}
\right]\varphi_1^4+\cdots, \label{V1HTE_MS}
\end{align}
where $\bar{\nu}_1^2(T)|_{\overline{\text{MS}}}=\bar{\nu}_1^2(T)|_{\Sigma_1=\Sigma_2=0}$.
The $\mathcal{O}(T^2)$ term in the first line breaks the RG invariance. 
After including terms arising from the sunset diagrams, they would become the RG invariant form, as shown below.

Taking $\mathcal{D}$ derivatives of $V_{\text{eff}}(\varphi_1)$ at the two-loop level, one finds
\begin{align}
\mathcal{D}V_0|_{\text{two-loop}} & = \beta_{\Omega}^{(2)}+\frac{\nu_1^2}{2}\beta_{\nu_1^2}^{(2)}\varphi_1^2+\frac{1}{4!}\beta_{\lambda_1}^{(2)}\varphi_1^4-(\nu_1^2+\Sigma_1)\gamma_{\Phi_1}^{(2)}\varphi_1^2-\frac{1}{3!}\gamma_{\Phi_1}^{(2)}\varphi_1^4 
\nonumber\\
& = -\frac{M_1^2\Sigma_1+M_2^2\Sigma_2}{16\pi^2}
-\frac{(\lambda_1^2+\lambda_3^2)M_1^2+2\lambda_3^2M_2^2}{2(16\pi^2)^2}\varphi_1^2-\Sigma_1\gamma_{\Phi_1}^{(2)}\varphi_1^2, \\
\mathcal{D}V_1|_{\text{two-loop}}
& = \frac{\bar{I}(M_1)}{2(16\pi^2)}\Big[\lambda_1M_1^2+\lambda_3M_2^2+(\lambda_1^2+\lambda_3^2)\varphi_1^2 \Big] \nonumber\\
&\quad +\frac{\bar{I}(M_2)}{2(16\pi^2)}\Big[\lambda_3M_1^2+\lambda_2M_2^2+2\lambda_3\varphi_1^2 \Big] +\Sigma_1\gamma_{\Phi_1}^{(2)}\varphi_1^2, \\
\mathcal{D}V_2|_{\text{two-loop}}
& = -\frac{\bar{I}(M_1)}{2(16\pi^2)}\Big[\lambda_1M_1^2+\lambda_3M_2^2+(\lambda_1^2+\lambda_3^2)\varphi_1^2 \Big] \nonumber\\
&\quad -\frac{\bar{I}(M_2)}{2(16\pi^2)}\Big[\lambda_3M_1^2+\lambda_2M_2^2+2\lambda_3\varphi_1^2 \Big] 
+\frac{(\lambda_1^2+\lambda_3^2)M_1^2+2\lambda_3^2M_2^2}{2(16\pi^2)^2}\varphi_1^2\nonumber\\
&\quad +\frac{M_1^2\Sigma_1+M_2^2\Sigma_2}{16\pi^2}.
\end{align}
Summing up, one gets $\mathcal{D}(V_0+V_1+V_2)|_{\text{two-loop}}=0$.

Let us look into what the $\bar{\mu}$-dependent terms look like using HTE.
\begin{align}
\lefteqn{V_{\text{eff}}^{\text{HTE}}(\varphi_1)=V_0(\varphi_1)+V_1^{\text{HTE}}(\varphi_1)+V_2^{\text{HTE}}(\varphi_1) }  \nonumber\\
&= \frac{1}{2}
\Bigg[
\bigg\{
\nu_1^2+\frac{\lambda_1\nu_1^2+\lambda_3\nu_2^2}{32\pi^2}\ln\frac{T^2}{\bar{\mu}^2} 
+\frac{2(\lambda_1^2+\lambda_3^2)(\nu_1^2+\Sigma_1)
	+\lambda_3(\lambda_1+\lambda_2+2\lambda_3)(\nu_2^2+\Sigma_2) }{4(16\pi^2)^2}\ln^2\frac{T^2}{\bar{\mu}^2} \nonumber\\
&\hspace{1.5cm}
-\frac{(\lambda_1^2+\lambda_3^2)(\nu_1^2+\Sigma_1)+2\lambda_3^2(\nu_2^2+\Sigma_2)}{2(16\pi^2)^2}\ln\frac{T^2}{\bar{\mu}^2}
\bigg\}
+ \frac{(\lambda_1^2+\lambda_2\lambda_3)T^2}{8(16\pi^2)}
\nonumber\\
&\hspace{1cm}
+\frac{T^2}{24}\left\{\left(\lambda_1+\frac{3\lambda_1^2+3\lambda_3^2}{32\pi^2}\ln\frac{T^2}{\bar{\mu}^2}\right)
+\left(\lambda_3+\frac{\lambda_3(\lambda_1+\lambda_2+4\lambda_3)}{32\pi^2}\ln\frac{T^2}{\bar{\mu}^2}\right) \right\} \nonumber\\
&\hspace{1cm}
+\bigg\{
\frac{\lambda_1(\nu_1^2+\Sigma_1)+\lambda_3(\nu_2^2+\Sigma_2) }{16\pi^2}
\nonumber\\
&\hspace{2cm}
+\frac{2(\lambda_1^2+\lambda_3^2)(\nu_1^2+\Sigma_1)
	+\lambda_3(\lambda_1+\lambda_2+2\lambda_3)(\nu_2^2+\Sigma_2) }{(16\pi^2)^2}\ln\frac{T^2}{\bar{\mu}^2}
\bigg\}c_B \nonumber\\
&\hspace{1cm}
+\frac{(\lambda_1^2+\lambda_3^2)(\nu_1^2+\Sigma_1)+\lambda_3(\lambda_1+\lambda_3)(\nu_2^2+\Sigma_2)}{(16\pi^2)^2}c_B^2
\Bigg]\varphi_1^2 \nonumber\\
&\quad
-\frac{T}{12\pi}
\bigg[
(M_1^2)^{3/2}+\frac{3}{4(16\pi^2)}\Big\{\lambda_1(M_1^2)^{3/2}+\lambda_3M_2^2(M_1^2)^{1/2}+ (\lambda_1^2+\lambda_3^2)(M_1^2)^{1/2}\varphi_1^2 \Big\}\ln\frac{T^2}{\bar{\mu}^2} \nonumber\\
&\hspace{1.5cm}
+(M_2^2)^{3/2}+\frac{3}{4(16\pi^2)}\Big\{\lambda_2(M_2^2)^{3/2}+\lambda_3M_1^2(M_2^2)^{1/2}+2\lambda_3^2(M_2^2)^{1/2}\varphi_1^2\Big\}\ln\frac{T^2}{\bar{\mu}^2} \nonumber\\
&\hspace{1.5cm}
+\frac{3}{2(16\pi^2)}\Big\{ \lambda_1(M_1^2)^{3/2}+\lambda_2(M_2^2)^{3/2}+\lambda_3\big(M_1^2(M_2^2)^{1/2}+M_2^2(M_1^2)^{1/2}\big) \Big\}c_B
\bigg] \nonumber\\
&\quad 
+\frac{1}{4!}
\Bigg[
\bigg\{
\lambda_1+\frac{3(\lambda_1^2+\lambda_3^2)}{32\pi^2}\ln\frac{T^2}{\bar{\mu}^2}
+\frac{3(3\lambda_1^3+4\lambda_1\lambda_3^2+\lambda_2\lambda_3^2+4\lambda_3^3)}{4(16\pi^2)}\ln^2\frac{T^2}{\bar{\mu}^2} \nonumber\\
&\hspace{2cm}
-\frac{3\left\{\lambda_1(\lambda_1^2+\lambda_3^2)+2\lambda_3^3\right\}}{(16\pi^2)^2}\ln\frac{T^2} {\bar{\mu}^2}
\bigg\} \nonumber\\
&\hspace{1.5cm}
+\frac{3}{16\pi^2}\left\{
\lambda_1^2+\lambda_3^2
+\frac{3\lambda_1^3+4\lambda_1\lambda_3^2+\lambda_2\lambda_3^2+4\lambda_3^3)}{16\pi^2}
\ln\frac{T^2}{\bar{\mu}^2}
\right\}c_B \nonumber\\
&\hspace{1.5cm}
+\frac{3(\lambda_1^3+\lambda_2\lambda_3^2+2\lambda_1\lambda_3^2)}{(16\pi^2)^2}c_B^2
\Bigg]\varphi_1^4+\frac{\lambda_3T^2}{4(16\pi^2)}(M_1^2)^{1/2}(M_2^2)^{1/2} + \cdots \nonumber\\
&= \frac{1}{2}
\bigg[\bar{\nu_1}^2(T)+\frac{T^2}{24}\left(\bar{\lambda}_1(T)+\bar{\lambda}_3(T)\right)
+\frac{1}{16\pi^2}\Big\{\bar{\lambda}_1(T)(\bar{\nu}_1^2(T)+\Sigma_1)+\bar{\lambda}_3(T)(\bar{\nu}_2^2(T)+\Sigma_2)\Big\}c_B \nonumber\\
&\hspace{1cm}
+\frac{(\lambda_1^2+\lambda_2\lambda_3)T^2}{8(16\pi^2)}
+\frac{(\lambda_1^2+\lambda_3^2)(\nu_1^2+\Sigma_1)+\lambda_3(\lambda_1+\lambda_3)(\nu_2^2+\Sigma_2)}{(16\pi^2)^2}c_B^2 
\bigg]\varphi_1^2 \nonumber\\
&\quad
-\frac{T}{12\pi}
\bigg[
\big(\bar{M}_1^2(T)\big)^{3/2}+\big(\bar{M}_2^2(T)\big)^{3/2} \nonumber\\
&\hspace{2cm}
+\frac{3}{2(16\pi^2)}\Big\{ \lambda_1(M_1^2)^{3/2}+\lambda_2(M_2^2)^{3/2}+\lambda_3\big(M_1^2(M_2^2)^{1/2}+M_2^2(M_1^2)^{1/2}\big) \Big\}c_B
\bigg] \nonumber\\
&\quad 
+\frac{1}{4!}
\left[
\bar{\lambda}_1(T)+\frac{3(\bar{\lambda}_1^2(T)+\bar{\lambda}_3^2(T))c_B}{16\pi^2}
+\frac{3(\lambda_1^3+\lambda_2\lambda_3^2+2\lambda_1\lambda_3^2)}{(16\pi^2)^2}c_B^2
\right]\varphi_1^4 \nonumber\\
&\quad  +\frac{\lambda_3T^2}{4(16\pi^2)}(M_1^2)^{1/2}(M_2^2)^{1/2}+\cdots.
\end{align}
Note that all the $\bar{\mu}$ dependencies are absorbed into the running parameters, and the RG invariance is manifest. 

\section{Gap equation and the consistency condition}\label{app:SigmaT}

\begin{figure}[H]
\begin{center}
\includegraphics[width=15cm]{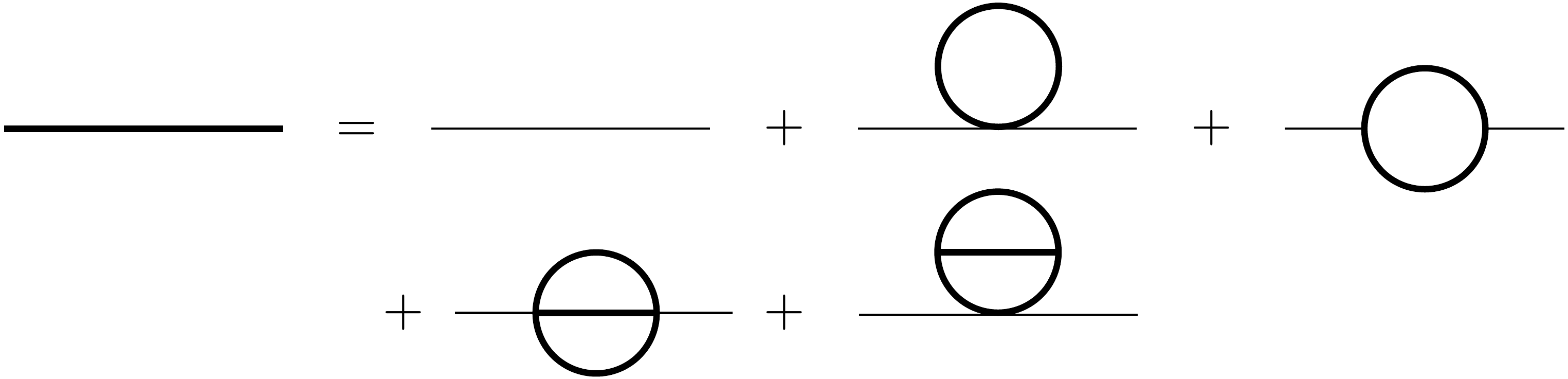}
\caption{The gap equation for the scalar mass. 
The thick solid lines represent the full propagators, while the thin lines are the tree ones.
The second diagram on the right-hand side incorporates daisy and super-daisy diagrams, 
which becomes dominant in the high-temperature limit~\cite{Dolan:1973qd}.}
\label{fig:gap_eq}
\end{center}
\end{figure}

We give more details about the consistency condition. For illustrative purposes, we focus on the $\phi^4$ theory.
After regularizing the gap equation in the $\overline{\text{MS}}$ scheme, it is cast into the form~\cite{Dolan:1973qd,Boyd:1993tz,Curtin:2016urg}
\begin{align}
M^2 & =\bar{m}^2+\Sigma(M^2)
 =\bar{m}^2+\frac{\bar{\lambda}}{2}\bar{I}(M)+\Sigma_\varphi(M^2)+\Sigma^{(\text{sunset})}(M^2),
\label{gapeq}
\end{align}
where the first three terms on the rightmost-hand side are pictorially represented as each diagram in the first line in Fig.~\ref{fig:gap_eq} 
and $\Sigma^{(\text{sunset})}$ includes the sunset-type diagrams shown in the second line in Fig.~\ref{fig:gap_eq}. 
We will use Eq.~(\ref{gapeq}) both at zero and finite temperatures to define $M^2$ and $M^2|_{T=0}$.
For now, we concentrate on the first three terms, where $\bar{I}(M)$ is defined as Eq.~(\ref{barI}) and $\Sigma_\varphi(M^2)$ as
\begin{align}
\Sigma_\varphi(M^2)& = \bar{\lambda}^2\bar{\varphi}^2\left[\frac{1}{2(16\pi^2)}\ln\frac{M^2}{\bar{\mu}_0^2}+\frac{1}{2\pi^2}I_B''(A^2) \right]-\frac{t\bar{\lambda}^2\bar{\varphi}^2}{16\pi^2} \nonumber \\
& \equiv \Sigma_\varphi(M^2)_{t=0}-\frac{t\bar{\lambda}^2\bar{\varphi}^2}{16\pi^2}.
\end{align}
In Eq.~(\ref{gapeq}), the $\overline{\text{MS}}$ running parameters are denoted as the barred quantities, and $I_B''(A^2\equiv M^2/T^2)=\partial^2 I_B(A^2)/\partial (A^2)^2$. Note that $\bar{I}(M)$ incorporates daisy and super-daisy diagrams but not sunset-type diagrams (see Ref.~\cite{Dolan:1973qd}). 
$\Sigma(M^2)$ without the sunset-type diagrams can be rearranged as
\begin{align}
\Sigma(M^2) 
& = -\frac{t\bar{\lambda}}{16\pi^2}(\bar{m}^2+\bar{\lambda}\bar{\varphi}^2)
+\frac{1}{2}
\left(
\bar{\lambda}-\frac{\bar{t\lambda}^2}{16\pi^2}
\right)\bar{I}(M)_{t=0}
+\left(1-\frac{t\bar{\lambda}}{16\pi^2}\right)\Sigma_\varphi(M^2)_{t=0} \nonumber\\
&\quad
+t^2\frac{\bar{\lambda}^2(M^2+\bar{\lambda}\bar{\varphi}^2)}{(16\pi^2)^2},
\end{align}
where we have used
\begin{align}
\bar{I}(M)& = \bar{I}(M)_{t=0}-\frac{2tM^2}{16\pi^2}= \bar{I}(M)_{t=0}-\frac{2t}{16\pi^2}\left[\bar{m}^2+\frac{\bar{\lambda}}{2}\bar{I}(M)+\Sigma_\varphi(M^2)\right] \nonumber\\
& = \bar{I}(M)_{t=0}-\frac{2t}{16\pi^2}\left[\bar{m}^2+\frac{\bar{\lambda}}{2}\left(\bar{I}(M)_{t=0}-\frac{2tM^2}{16\pi^2} \right) +\Sigma_\varphi(M^2)_{t=0}-\frac{t\bar{\lambda}^2\bar{\varphi}^2}{16\pi^2} \right].
\end{align}
The thermal correction $\Sigma(T)$ that is used in Eq.(\ref{LB}) is defined as
\begin{align}
\Sigma(T) &\equiv M^2-M^2|_{T=0} \nonumber\\
& = \frac{1}{2}
\left(
\bar{\lambda}-\frac{\bar{t\lambda}^2}{16\pi^2}
\right)\left(\bar{I}(M)_{t=0}-\bar{I}_0(M)_{t=0}|_{T=0} \right) \nonumber\\
&\quad
	+\left(1-\frac{t\bar{\lambda}}{16\pi^2}\right)\Big(\Sigma_\varphi(M^2)_{t=0}-\Sigma_\varphi(M^2)_{t=0}|_{T=0}\Big) +\frac{t^2\bar{\lambda}^2\Sigma(T)}{(16\pi^2)^2} \nonumber\\
& \simeq \frac{T^2}{2\pi^2}
\left(
\bar{\lambda}-\frac{t\bar{\lambda}^2}{16\pi^2}
\right)I'_B(A^2)+\frac{\bar{\varphi}^2}{2\pi^2}\left(\bar{\lambda}^2-\frac{t\bar{\lambda}^3}{16\pi^2}\right)I_B''(A^2)+\cdots,
\label{SigmaT}
\end{align}
where the ellipsis in the last line denotes higher-order terms that are irrelevant below.
Now, we show the $t$ independence of $M^2$ and $\Sigma(T)$. Taking the $t$ derivative of the gap equation (\ref{gapeq}), one has
\begin{align}
\frac{dM^2}{dt} &= \frac{d\bar{m}^2}{dt}+\frac{d\Sigma(M^2)|_{T=0}}{dt}+\frac{d\Sigma(T)}{dt}.
\end{align}
At the one-loop order, the zero-temperature part is calculated as
\begin{align}
\frac{d\bar{m}^2}{dt}+\frac{d\Sigma(M^2)|_{T=0}}{dt}
&= \frac{\bar{\lambda}}{16\pi^2}(\bar{m}^2+\bar{\lambda}\bar{\varphi}^2)-\frac{\bar{\lambda}}{16\pi^2}(\bar{m}^2+\bar{\lambda}\bar{\varphi}^2)= 0.
\end{align}
For the temperature-dependent part $\Sigma(T)$, on the other hand, it follows from Eq.~(\ref{SigmaT}) that 
\begin{align}
\frac{d\Sigma(T)}{dt}&=
\frac{T^2}{2\pi^2}
\left(
	\beta_\lambda^{(1)}-\frac{\bar{\lambda}^2}{16\pi^2} 
\right)I'_B(A^2)
+\frac{\bar{\varphi}^2}{2\pi^2}\left(2\bar{\lambda}\beta_\lambda^{(1)}-\frac{\bar{\lambda}^3}{16\pi^2}\right)I_B''(A^2)+\cdots.
\end{align}
Since this quantity is the two-loop order, it can be ignored in the calculation of the one-loop order.

Now we move on to the two-loop order calculation focusing on $\Sigma(T)$. 
In this case, the contributions of the sunset-type diagrams are necessary to show the $t$ independence of $\Sigma(T)$. 
It is straightforward to derive the two-loop contribution by taking the second derivative of $V_2$ with respect to $\bar{\varphi}$ and replacing $\bar{m}^2$ with $M^2$.
The relevant part has the form
\begin{align}
\frac{\partial^2 V_2}{\partial \bar{\varphi}^2}\bigg|_{\bar{m}^2\to M^2}
&\ni \Sigma^{(\text{sunset})}(M^2) = 
\left[
\frac{\bar{\lambda}^2T^2}{2\pi^2(16\pi^2)}I_B'(A^2)
+\frac{5\bar{\lambda}^3\bar{\varphi}^2}{4\pi^2(16\pi^2)}I_B''(A^2)
\right]\ln\frac{M^2}{\bar{\mu}_0^2e^{2t}}.
\end{align}
After taking the two-loop contributions into account, $\Sigma(T)$ is modified to
\begin{align}
\Sigma(T) & \simeq \frac{T^2}{2\pi^2}\left(\bar{\lambda}-t\frac{3\bar{\lambda}^2}{16\pi^2}\right)I_B'(A^2)
+\frac{\bar{\varphi}^2}{2\pi^2}\left(\bar{\lambda}^2-t\frac{6\bar{\lambda}^3}{16\pi^2}\right)I_B''(A^2),
\label{2LSigmaT}
\end{align}
from which, since $\beta_\lambda^{(1)} = 3\bar{\lambda}^2/16\pi^2$, it follows that
\begin{align}
\frac{d\Sigma(T)}{dt} &= \frac{T^2}{2\pi^2}\left(\beta_\lambda^{(1)}-\frac{3\bar{\lambda}^2}{16\pi^2}\right)I_B'(A^2)
+\frac{\bar{\varphi}^2}{2\pi^2}\left(2\bar{\lambda}\beta_\lambda^{(1)}-\frac{6\bar{\lambda}^3}{16\pi^2}\right)I_B''(A^2)
+\mathcal{O}\left(\frac{\bar{\lambda}^3}{(16\pi^2)^3}\right) \nonumber\\
& =0+\mathcal{O}\left(\frac{\bar{\lambda}^3}{(16\pi^2)^3}\right).
\label{dSigmadt_2L}
\end{align}
Therefore, $\Sigma(T)$ is $t$ independent up to the two-loop order. 
We emphasize that, besides the $\overline{\text{MS}}$ scheme, it is possible to formulate a scheme such that $d\Sigma(T)/dt=0$ by modifying the $\beta$-functions, although in a nontrivial form.
In any scheme, the $t$-invariant $\Sigma(T)$ up to a certain order can be obtained, which underpins our consistency condition.

In the high-temperature limit with $\bar{\lambda}/24< 1$, Eq.~(\ref{2LSigmaT}) is reduced to
\begin{align}
\Sigma(T)\simeq \frac{\bar{\lambda}T^2}{24}
\left[1-t\frac{3\bar{\lambda}}{16\pi^2}+\mathcal{O}\left(\frac{\bar{\varphi}^2}{T^2}\right)\right].
\end{align}
The $t$ invariance of $\Sigma(T)$ allows one to have the simplified form $\Sigma(T)=\lambda T^2/24$ by choosing $t=0$, and we use it in this work. 
By the same token, $\Sigma_1(T)$ and $\Sigma_2(T)$ in the extension of the $\phi^4$ model discussed in Sec.~\ref{sec:Exphi4} are also defined at $t=0$.

\section{Loop functions}\label{app:therfuncs}
Let us define the sum-integral symbol as
\begin{align}
\sumint_k \equiv \mu^\epsilon T\sum_{n=-\infty}^\infty\int\frac{d^{d-1}\boldsymbol{k}}{(2\pi)^{d-1}},
\end{align}
where $n$ denote integers and $d=4-\epsilon$. 
A thermal function for the one-loop bubble diagram is defined as
\begin{align}
I(m) &= \sumint_k \frac{1}{k^2+m^2}
=-\frac{m^2}{16\pi^2}\frac{2}{\epsilon}+\bar{I}(m)
	+\epsilon i_\epsilon(m)+\mathcal{O}(\epsilon^2), 
\end{align}
with 
\begin{align}
\bar{I}(m) &= \frac{m^2}{16\pi^2}\left(\ln\frac{m^2}{\bar{\mu}^2}-1\right)
	+\frac{T^2}{\pi^2}I'_{B}\left(\frac{m^2}{T^2}\right) \equiv \bar{I}_0(m)+\frac{T^2}{\pi^2}I'_{B}\left(\frac{m^2}{T^2}\right),
\label{barI}
\end{align}
where $k^2=\omega_n^2+\boldsymbol{k}^2$ with $\omega_n=2n\pi T$.\footnote{We focus exclusively on the bosonic case.} 
The explicit form of $i_\epsilon(m)$, which is needed when one goes beyond the one-loop level, is 
\begin{align}
i_\epsilon(m)&= 
	-\frac{m^2}{64\pi^2}\bigg[\bigg(\ln\frac{m^2}{\bar{\mu}^2}-1\bigg)^2
	+1+\frac{\pi^2}{6}\bigg]
	-\frac{T^2}{2\pi^2}\bigg[\bigg(\ln\frac{T^2}{\bar{\mu}^2}+\ln4-2\bigg)I'_{B}(a^2) + j(a^2)\bigg],
\end{align}
where
\begin{align}
j(a^2) &= \int^\infty_0dx
	\frac{x^2\ln x}{\sqrt{x^2+a^2}}\frac{1}{e^{\sqrt{x^2+a^2}}-1}.
\end{align}
The contributions from the function $i_\epsilon(m)$ are cancelled among the diagrams and do not appear in the renormalized effective potential.

The sunset-type diagram composed of all the scalar fields is defined as
\begin{align}
H(m_1,m_2,m_3)
&= \sumint_k\sumint_q\frac{1}{(k^2+m^2_1)(q^2+m^2_2)[(k+q)^2+m^2_3]}, 
\end{align}
where $k^2=\omega_n^2+\boldsymbol{k}^2$ and $q^2=\omega_m^2+\boldsymbol{q}^2$ with $\omega_n^2=2n\pi T$ and $\omega_m^2=2m\pi T$.
We parametrize $H(m_1, m_2, m_3)$ in terms of the divergent and finite parts as
\begin{align}
H(m_1, m_2, m_3) &= H^{\rm div}(m_1, m_2, m_3) +\tilde{H}(m_1, m_2, m_3) 
	+\frac{1}{8\pi^2}\sum_{j=1}^3i_\epsilon(m_j) 
\end{align}
where
\begin{align}
H^{\rm div}(m_1, m_2, m_3) 
&= -\frac{1}{(16\pi^2)^2}\left(\frac{2}{\epsilon^2}
	+\frac{1}{\epsilon}\right)(m^2_1+m^2_2+m^2_3)
	+\frac{1}{16\pi^2}\frac{2}{\epsilon}\Big(\bar{I}(m_1)+\bar{I}(m_2)+\bar{I}(m_3)\Big).
\end{align}
The divergences in the first line are removed by the local counterterms. As discussed in Sec.~\ref{sec:beta}, only a single $\epsilon$ pole contributes to the $\beta$-functions.
On the other hand, the divergences proportional to $\bar{I}(m)$ are cancelled among the diagrams.

The finite part is given by
\begin{align}
\lefteqn{\tilde{H}(m_1,m_2,m_3)} \nonumber\\
&= \frac{1}{16\pi^2}\big(\bar{I}_0(m_1)+\bar{I}_0(m_2)+\bar{I}_0(m_3) \big)
	-\frac{1}{(4\pi)^4}(m_1^2+m_2^2+m_3^2) \nonumber\\
&\quad -\frac{1}{2}
	\Big\{
		\frac{m_1^2+m_2^2-m_3^2}{m_1^2m_2^2}\bar{I}_0(m_1)\bar{I}_0(m_2)
		+\frac{m_2^2+m_3^2-m_1^2}{m_2^2m_3^2}\bar{I}_0(m_2)\bar{I}_0(m_3) \nonumber\\
&\hspace{2cm}
		+\frac{m_3^2+m_1^2-m_2^2}{m_3^2m_1^2}\bar{I}_0(m_3)\bar{I}_0(m_1)		
	\Big\} +\frac{1}{(4\pi)^4}R{\it \Phi}(m_1, m_2, m_3) \nonumber\\
&\quad -\frac{T^2}{(2\pi)^4}
\Big[
	\varphi(m_1, m_2, m_3)I'_B(a_1^2)
	+\varphi(m_2, m_3, m_1)I'_B(a_2^2) 
	+\varphi(m_3, m_1, m_2)I'_B(a_3^2)
\Big] \nonumber\\
&\quad+\frac{T^2}{4(2\pi)^4} 
	\Big[K_{--}(a_1, a_2, a_3)+K_{--}(a_2, a_3, a_1)+K_{--}(a_3, a_1, a_2)\Big].
\label{tilH123}
\end{align}
where $R^2= (m^2_1+m^2_2-m^2_3)^2-4m^2_1m^2_2$ and
\begin{align}
{\it \Phi}(m_1,m_2,m_3)&= {\rm Li}_2\left(\frac{m^2_1+m^2_2-m^2_3-R}{2m^2_1}\right)
	+{\rm Li}_2\left(\frac{m^2_1-m^2_2+m^2_3-R}{2m^2_1}\right) 
	+\frac{1}{2}\ln\frac{m^2_2}{m^2_1}\ln\frac{m^2_3}{m^2_1}\nonumber\\
&\quad-\ln\left(\frac{m^2_1+m^2_2-m^2_3-R}{2m^2_1}\right)\ln\left(\frac{m^2_1-m^2_2+m^2_3-R}{2m^2_1}\right)
	-\frac{\pi^2}{6}.
\end{align}
Note that the dilogarithmic function ${\rm Li}_2(z)$ has an imaginary part if $z>1$, i.e., if $m_1^2-m_2^2+m_3^2+R<0$ or  $m_1^2+m_2^2-m_3^2+R<0$, ${\rm Li}_2$ in the first line has the imaginary part. However, the log term in the second line also has the imaginary part that cancels the imaginary part of the former. For the numerical calculation of $\mathit{\Phi}(m_1, m_2, m_3)$, to evaluate the real part, we use
\begin{align}
{\rm Re}\big[{\rm Li}_2(z)\big] = \frac{\pi^2}{6}-\int^z_1dt~\frac{\ln|1-t|}{t}.
\end{align}
Furthermore, for $R^2<0$, $R$ and $\mathit{\Phi}(m_1, m_2, m_3)$ have the imaginary parts. They are cancelled to each other and $R\mathit{\Phi}(m_1, m_2, m_3)$ is reduced to
\begin{align}
R\mathit{\Phi}(m_1, m_2, m_3) &= |R|
	\left[
	2\int^1_0\frac{dt}{t}~\tan^{-1}
		\left(
		\frac{m_2t\sin\eta}{m_1-m_2t\cos\eta}
		\right)
		+\theta \ln\frac{m^2_2}{m^2_1}
	\right],
\end{align}
where
\begin{align}
\eta 
=\arctan\left(\frac{|R|}{m^2_1+m^2_2-m^2_3}\right), \quad
\theta = \arctan\left(\frac{-|R|}{m^2_1-m^2_2+m^2_3}\right),
\end{align}
and
$\varphi(m_1, m_2, m_3)$ is defined as
\begin{align}
\varphi(m_1, m_2, m_3) = \int^1_0dx~\ln
	\left(
	\frac{-x(1-x)m^2_1+(1-x)m^2_2+xm^2_3}{\bar{\mu}^2}
	\right),
\end{align}
and 
\begin{align}
K_{--}(a_1, a_2, a_3)&= \int^\infty_0dx~\frac{x n_-(x; a_1)}{\sqrt{x^2+a^2_1}}
	\int^\infty_0dy~\frac{y n_-(y; a_2)}{\sqrt{y^2+a^2_2}}
	\ln\left|
	\frac{\tilde{Y}_+(x, y; a_1, a_2, a_3)}{\tilde{Y}_-(x, y; a_1, a_2, a_3)}
	\right|
\end{align}
with 
\begin{align}
n_-(x;a) &= \frac{1}{e^{\sqrt{x^2+a^2}}-1}, \\
\tilde{Y}_\pm(x, y; a_1, a_2, a_3)
&=
\Big[
	(a_1^2+a_2^2-a_3^2)^2-4a_1^2a_2^2-4\big\{a_2^2x^2\pm(a_1^2+a_2^2-a_3^2)xy+a_1^2y^2\big\}
\Big]^2.
\end{align}
For $m_1=m_2=m_3$, $\tilde{H}(m, m, m)$ is reduced to 
\begin{align}
\tilde{H}(m)&\equiv \tilde{H}(m, m, m)
= 3
\bigg[
	-\frac{\bar{I}^2(m)}{2m^2}+\frac{\bar{I}(m)}{16\pi^2}-\frac{m^2}{(16\pi^2)^2}\left(1+\frac{2}{3}f_2\right)
	 \nonumber\\
&\hspace{3.5cm}	
	-\frac{1}{2m^2}\frac{T^2}{\pi^2}\big(I_B'(a^2)\big)^2-\frac{T^2}{16\sqrt{3}\pi^3}I_B'(a^2) +\frac{4T^2}{(16\pi^2)^2}K(a)
\bigg], \label{tilH}
\end{align}
where $K(a)\equiv K_{--}(a,a,a)$ and we have used 
\begin{align}
\varphi(m, m, m)&= \ln\frac{m^2}{\bar{\mu}^2}-2+\frac{\pi}{\sqrt{3}}, \\
\it{\Phi}(m,m,m) &=-\frac{\pi^2}{18}+2{\rm Li}_2\left(\frac{1-\sqrt{3}i}{2}\right), 
\end{align}
and $f_2=-\frac{\sqrt{3}}{2}i{\it\Phi}(m,m,m)\simeq-1.76$.

In our numerical analysis, we use an approximation~\cite{Arnold:1992rz}
\begin{align}
K_{--}(a_1 ,a_2, a_3)=K\left(\frac{a_1+a_2+a_3}{3}\right).
\end{align}

%
%
\section{Tadpole and mass conditions for RG-improved one-loop effective potentials}
Some parameters in the Lagrangian can be expressed in terms of VEV and the scalar masses using tadpole and mass conditions. 
In the cases of the RG-improved effective potentials with our $t(\varphi)$, their relations are more involved than those in fixed-order calculations. In this Appendix, we explicitly give the first and second derivatives of the RG-improved one-loop effective potentials with respect to the background fields.
Although we do not use such a potential in our numerical analysis in the $\phi^4$ theory, we still present all the formulas to know how they differ from the fixed-order expressions.

\subsection{$\phi^4$ theory}
At zero temperature, the $t$-$\varphi$ relation (\ref{t-phi}) is reduced to $t(\varphi)=\ln(\bar{m}^2/e\bar{\mu}_0^2)/2$. With this, the one-loop effective potential is cast into the form
\begin{align}
\bar{V}_{\text{eff}}(\varphi; t(\varphi))=\bar{V}_0(\varphi;t(\varphi))+\bar{V}_1(\varphi;t(\varphi)) =\bar{\Omega}-\frac{\bar{\nu}^2}{2}\varphi^2+\frac{\bar{\lambda}}{4!}\varphi^4
-\frac{\bar{m}^4}{8(16\pi^2)},
\end{align}
where $\bar{m}^2=-\bar{\nu}^2+\bar{\lambda}\varphi^2/2$. The first derivative of $\bar{V}_{\text{eff}}(\varphi; t(\varphi))$ with respect to $\varphi$ is
\begin{align}
\frac{d \bar{V}_{\text{eff}}(\varphi; t(\varphi))}{d \varphi} &= \frac{\partial \bar{V}_{\text{eff}}(\varphi; t(\varphi))}{\partial \varphi} 
+\frac{d t(\varphi)}{d\varphi} \frac{\partial \bar{V}_{\text{eff}}(\varphi; t)}{\partial t}\bigg|_{t = t(\varphi)} \nonumber\\
&= \varphi\left[-\bar{\nu}^2+\frac{\bar{\lambda}}{6}\varphi^2-\frac{\bar{\lambda}\bar{m}^2}{4(16\pi^2)}\right]
+\frac{d t(\varphi)}{d\varphi}\cdot \frac{\bar{m}^2(2\bar{m}^2-\mathcal{N})}{4(16\pi^2)} \nonumber\\
& =\varphi\left[-\bar{\nu}^2+\frac{\bar{\lambda}}{6}\varphi^2\right],
\end{align}
where $\mathcal{N} = \bar{\lambda}\left(\bar{m}^2+\bar{\lambda}\varphi^2\right)/16\pi^2$ and
\begin{align}
\frac{dt(\varphi)}{d\varphi} =\frac{\bar{\lambda}\varphi}{2\bar{m}^2-\mathcal{N}}.
\end{align}
Since we determine $\bar{\mu}_0$ by the condition $t(\varphi=v)=0$, i.e., $\bar{\mu}_0^2 = (-\nu^2+\lambda v^2/2)/e$, it is easy to solve the tadpole condition $(d\bar{V}_{\text{eff}}/d\varphi)|_{\varphi=v}=0$, which gives
\begin{align}
\nu^2 = \frac{\lambda}{6}v^2.
\end{align}
The second derivative of $\bar{V}_{\text{eff}}(\varphi; t(\varphi))$ is found to be
\begin{align}
\frac{d^2 \bar{V}_{\text{eff}}(\varphi; t(\varphi))}{d \varphi^2}
 & = -\bar{\nu}^2+\frac{\bar{\lambda}}{2}\varphi^2+\frac{\bar{\lambda}^2\varphi^2}{2(16\pi^2)}\frac{1}{1-\mathcal{N}/2\bar{m}^2}.
\end{align}
Thus, the mass in the vacuum (denoted as $m_\phi$) is obtained by
\begin{align}
m_\phi^2 = \frac{d^2 \bar{V}_{\text{eff}}(\varphi; t(\varphi))}{d \varphi^2}\bigg|_{\varphi=v}
 & = \frac{\lambda}{3}v^2\frac{1-\lambda/32\pi^2}{1-\lambda/8\pi^2} \nonumber\\
 & = \frac{\lambda}{3}v^2\left[1+\frac{3\lambda}{2(16\pi^2)}+\frac{3\lambda^2}{(16\pi^2)^2}+\cdots \right].
\end{align}
One should note that $m_\phi$ would agree with a one-loop fixed-order result if the higher-order terms are dropped, as it should.
\subsection{$\phi^4$ theory with an additional real scalar}
We obtain the first and second derivatives of $\bar{V}_{\text{eff}}(\varphi_1;t(\varphi_1))$ in Eq.~(\ref{barVeff1L_Exphi4}) with the $t$-$\varphi_1$ relation (\ref{t-phi_Exphi4}) at zero temperature. The first derivative of $\bar{V}_{\text{eff}}$ with respect to $\varphi_1$ is 
\begin{align}
\frac{d \bar{V}_{\text{eff}}(\bar{\varphi}_1; t(\varphi_1))}{d \varphi_1} 
& = \varphi_1
\left[
	\bar{\nu}_1^2+\frac{\bar{\lambda}_1}{6}\varphi_1^2
	+\frac{1}{2}\left(\bar{\lambda}_1\bar{I}_0(\bar{m}_1)+\bar{\lambda}_3\bar{I}_0(\bar{m}_2)\right)
\right]=0,
\end{align}
where
\begin{align}
\bar{I}_0(\bar{m}) = \frac{\bar{m}^2}{16\pi^2}\left(\ln\frac{\bar{m}^2}{e^{2t}\bar{\mu}_0^2}-1\right).
\end{align}
As in the $\phi^4$ theory, we determine $\bar{\mu}_0$ by the condition $\frac{\partial \bar{V}_{\text{eff}}(\bar{\varphi}_1; t)}{\partial t}|_{t=0} =0$, i.e., $t(\varphi_1=v)=0$, from which it follows that 
\begin{align}
\ln \bar{\mu}_0^2
& = \frac{\sum_{i=1,2}\frac{\partial \bar{m}_i^2}{\partial t}|_{t=0}m_i^2(\ln m_i^2-1)}{\sum_{i=1,2}\frac{\partial \bar{m}_i^2}{\partial t}|_{t=0}m_i^2},\label{mu0}
\end{align}
where
\begin{align}
m_1^2 &= \nu_1^2+\frac{\lambda_1}{2}v^2,\quad m_2^2 = \nu_2^2+\frac{\lambda_3}{2}v^2, \\
\frac{\partial \bar{m}_1^2}{\partial t}\bigg|_{t=0} 
& = \frac{1}{16\pi^2}\left[\lambda_1\nu_1^2+\lambda_3\nu_2^2+\frac{3}{2}(\lambda_1^2+\lambda_3^2)v^2\right], \\
\frac{\partial \bar{m}_2^2}{\partial t}\bigg|_{t=0} 
& = \frac{1}{16\pi^2}\left[\lambda_3\nu_1^2+\lambda_2\nu_2^2+\frac{1}{2}\lambda_3(\lambda_1+\lambda_2+4\lambda_3)v^2\right].
\end{align}
With this $\bar{\mu}_0$, the tadpole condition is simplified to
\begin{align}
\frac{d \bar{V}_{\text{eff}}(\bar{\varphi}_1; t(\varphi_1))}{d \varphi_1} \bigg|_{\varphi_1=v}
 = v
\left[
	\nu_1^2+\frac{\lambda_1}{6}v^2
	+\frac{1}{2}\left(\lambda_1\bar{I}_0(m_1)+\lambda_3\bar{I}_0(m_2)\right)
\right]=0,
\end{align}
which determines $\nu_1^2$ as
\begin{align}
\nu_1^2 = -
\left[\frac{\lambda_1}{6}v^2
	+\frac{\lambda_1m_1^2}{32\pi^2}\left(\ln\frac{m_1^2}{\bar{\mu}_0^2}-1\right)
	+\frac{\lambda_3m_2^2}{32\pi^2}\left(\ln\frac{m_2^2}{\bar{\mu}_0^2}-1\right)
\right].
\end{align}
This coincides with the one-loop fixed-order result, but $\bar{\mu}_0$ is given by Eq.~(\ref{mu0}).

The second derivative is cast into the form
\begin{align}
m_{\phi_1}^2 & = \frac{d^2 \bar{V}_{\text{eff}}(\varphi_1; t(\varphi_1))}{d \varphi_1^2}\bigg|_{\varphi_1 = v} \nonumber\\
&  =  m_1^2
 +\frac{1}{2}
 \Big[
 	\lambda_1\bar{I}_0(m_1)+\lambda_3\bar{I}_0(m_2)+\left( \lambda_1^2\bar{I}_0'(m_1)+\lambda_3^2\bar{I}_0'(m_2) \right)v^2
 \Big] \nonumber\\
 &\quad +\frac{dt(\varphi_1)}{d\varphi_1}\bigg|_{\varphi_1 = v}\frac{1}{2}\sum_{i=1,2}
 \left[
 	\frac{\partial^2 \bar{m}_i^2}{\partial\varphi_1\partial t}\bar{I}_0(m_i)
	+\frac{\partial \bar{m}_i^2}{\partial\varphi_1}\frac{\partial \bar{m}_i^2}{\partial t}\bar{I}_0'(m_i)
 \right]_{t=0},
\end{align}
where
\begin{align}
\frac{dt(\varphi_1)}{d\varphi_1}\bigg|_{\varphi_1=v}
 = \frac{  \sum_{i=1,2}
 \left[
 	\frac{\partial^2 \bar{m}_i^2}{\partial\varphi_1\partial t}\bar{I}_0(m_i)
	+\frac{\partial \bar{m}_i^2}{\partial\varphi_1}\frac{\partial \bar{m}_i^2}{\partial t}\bar{I}'_0(m_i)
 \right]_{t=0} }
 { \sum_{i=1,2}
 \left[
	 \frac{m_i^2}{8\pi^2}\frac{\partial \bar{m}_i^2}{\partial t}
 	-\frac{\partial^2 \bar{m}_i^2}{\partial t^2}\bar{I}_0(m_i)
	-\left(\frac{\partial^2 \bar{m}_2^2}{\partial t^2}\right)^2\bar{I}'_0(m_i)
 \right]_{t=0} },
\end{align}
and
\begin{align}
\frac{\partial^2 \bar{m}_1^2}{\partial\varphi_1\partial t}
 &= \frac{3(\bar{\lambda}_1^2+\bar{\lambda}_3^2)\varphi_1}{16\pi^2},\\
\frac{\partial^2 \bar{m}_2^2}{\partial\varphi_1\partial t}
 &= \frac{\bar{\lambda}_3(\bar{\lambda}_1+\bar{\lambda}_2+4\bar{\lambda}_3)\varphi_1}{16\pi^2},\\
 \frac{\partial^2 \bar{m}_1^2}{\partial t^2}
& = \frac{1}{16\pi^2}
\left[
	\beta_{\lambda_1}^{(1)}(\bar{m}_1^2+2\bar{\lambda}_1\varphi_1^2)
	+\beta_{\lambda_3}^{(1)}(\bar{m}_2^2+2\bar{\lambda}_3\varphi_1^2)
	+\bar{\lambda}_1\frac{\partial \bar{m}_1^2}{\partial t}
	+\bar{\lambda}_3\frac{\partial \bar{m}_2^2}{\partial t}
\right] \nonumber\\
& = \frac{1}{(16\pi^2)^2}
\Big[
	4(\bar{\lambda}_1^2+\bar{\lambda}_3^2)\bar{m}_1^2
	+2\bar{\lambda}_3(\bar{\lambda}_1+\bar{\lambda}_2+2\bar{\lambda}_3)\bar{m}_2^2 \nonumber\\
&\hspace{2cm}
	+\Big\{
		7\bar{\lambda}_1(\bar{\lambda}_1^2+\bar{\lambda}_3^2)
		+2\bar{\lambda}_3^2(\bar{\lambda}_1+\bar{\lambda}_2+5\bar{\lambda}_3)
	\Big\}\varphi_1^2
\Big], \\
\frac{\partial^2 \bar{m}_2^2}{\partial t^2}
& = \frac{1}{16\pi^2}
\left[
	\beta_{\lambda_3}^{(1)}(\bar{m}_1^2+4\bar{\lambda}_3\varphi_1^2)+\beta_{\lambda_2}^{(1)}\bar{m}_2^2
	+\bar{\lambda}_3\frac{\partial \bar{m}_1^2}{\partial t}
	+\bar{\lambda}_2\frac{\partial \bar{m}_2^2}{\partial t}
\right] \nonumber\\
& = \frac{1}{(16\pi^2)^2}
\Big[
	2\bar{\lambda}_3(\bar{\lambda}_1+\bar{\lambda}_2+2\bar{\lambda}_3)\bar{m}_1^2
	+4(\bar{\lambda}_2^2+\bar{\lambda}_3^2)\bar{m}_2^2 \nonumber\\
&\hspace{2.5cm}
	+\bar{\lambda}_3(
		\bar{\lambda}_1^2+4\bar{\lambda}_1\bar{\lambda}_3+6\bar{\lambda}_2\bar{\lambda}_3
		+17\bar{\lambda}_3^2
		)\varphi_1^2
\Big], \\
\bar{I}'_0(m)&=\frac{d \bar{I}_0(m)}{d m^2} = \frac{1}{16\pi^2}\ln\frac{m^2}{e^{2t}\bar{\mu}_0^2}.
\end{align}

%
\bibliography{refs}

\begin{thebibliography}{40}%
\makeatletter
\providecommand \@ifxundefined [1]{%
 \@ifx{#1\undefined}
}%
\providecommand \@ifnum [1]{%
 \ifnum #1\expandafter \@firstoftwo
 \else \expandafter \@secondoftwo
 \fi
}%
\providecommand \@ifx [1]{%
 \ifx #1\expandafter \@firstoftwo
 \else \expandafter \@secondoftwo
 \fi
}%
\providecommand \natexlab [1]{#1}%
\providecommand \enquote  [1]{``#1''}%
\providecommand \bibnamefont  [1]{#1}%
\providecommand \bibfnamefont [1]{#1}%
\providecommand \citenamefont [1]{#1}%
\providecommand \href@noop [0]{\@secondoftwo}%
\providecommand \href [0]{\begingroup \@sanitize@url \@href}%
\providecommand \@href[1]{\@@startlink{#1}\@@href}%
\providecommand \@@href[1]{\endgroup#1\@@endlink}%
\providecommand \@sanitize@url [0]{\catcode `\\12\catcode `\$12\catcode
  `\&12\catcode `\#12\catcode `\^12\catcode `\_12\catcode `\%12\relax}%
\providecommand \@@startlink[1]{}%
\providecommand \@@endlink[0]{}%
\providecommand \url  [0]{\begingroup\@sanitize@url \@url }%
\providecommand \@url [1]{\endgroup\@href {#1}{\urlprefix }}%
\providecommand \urlprefix  [0]{URL }%
\providecommand \Eprint [0]{\href }%
\providecommand \doibase [0]{http://dx.doi.org/}%
\providecommand \selectlanguage [0]{\@gobble}%
\providecommand \bibinfo  [0]{\@secondoftwo}%
\providecommand \bibfield  [0]{\@secondoftwo}%
\providecommand \translation [1]{[#1]}%
\providecommand \BibitemOpen [0]{}%
\providecommand \bibitemStop [0]{}%
\providecommand \bibitemNoStop [0]{.\EOS\space}%
\providecommand \EOS [0]{\spacefactor3000\relax}%
\providecommand \BibitemShut  [1]{\csname bibitem#1\endcsname}%
\let\auto@bib@innerbib\@empty
\bibitem [{\citenamefont {Rubakov}\ and\ \citenamefont
  {Shaposhnikov}(1996)}]{Rubakov:1996vz}%
  \BibitemOpen
  \bibfield  {author} {\bibinfo {author} {\bibfnamefont {V.~A.}\ \bibnamefont
  {Rubakov}}\ and\ \bibinfo {author} {\bibfnamefont {M.~E.}\ \bibnamefont
  {Shaposhnikov}},\ }\href {\doibase 10.1070/PU1996v039n05ABEH000145}
  {\bibfield  {journal} {\bibinfo  {journal} {Usp. Fiz. Nauk}\ }\textbf
  {\bibinfo {volume} {166}},\ \bibinfo {pages} {493} (\bibinfo {year}
  {1996})},\ \bibinfo {note} {[Phys. Usp.39,461(1996)]},\ \Eprint
  {http://arxiv.org/abs/hep-ph/9603208} {arXiv:hep-ph/9603208 [hep-ph]}
  \BibitemShut {NoStop}%
\bibitem [{\citenamefont {Funakubo}(1996)}]{Funakubo:1996dw}%
  \BibitemOpen
  \bibfield  {author} {\bibinfo {author} {\bibfnamefont {K.}~\bibnamefont
  {Funakubo}},\ }\href {\doibase 10.1143/PTP.96.475} {\bibfield  {journal}
  {\bibinfo  {journal} {Prog. Theor. Phys.}\ }\textbf {\bibinfo {volume}
  {96}},\ \bibinfo {pages} {475} (\bibinfo {year} {1996})},\ \Eprint
  {http://arxiv.org/abs/hep-ph/9608358} {arXiv:hep-ph/9608358 [hep-ph]}
  \BibitemShut {NoStop}%
\bibitem [{\citenamefont {Riotto}(1998)}]{Riotto:1998bt}%
  \BibitemOpen
  \bibfield  {author} {\bibinfo {author} {\bibfnamefont {A.}~\bibnamefont
  {Riotto}},\ }in\ \href@noop {} {\emph {\bibinfo {booktitle} {{Proceedings,
  Summer School in High-energy physics and cosmology: Trieste, Italy, June
  29-July 17, 1998}}}}\ (\bibinfo {year} {1998})\ pp.\ \bibinfo {pages}
  {326--436},\ \Eprint {http://arxiv.org/abs/hep-ph/9807454}
  {arXiv:hep-ph/9807454 [hep-ph]} \BibitemShut {NoStop}%
\bibitem [{\citenamefont {Trodden}(1999)}]{Trodden:1998ym}%
  \BibitemOpen
  \bibfield  {author} {\bibinfo {author} {\bibfnamefont {M.}~\bibnamefont
  {Trodden}},\ }\href {\doibase 10.1103/RevModPhys.71.1463} {\bibfield
  {journal} {\bibinfo  {journal} {Rev. Mod. Phys.}\ }\textbf {\bibinfo {volume}
  {71}},\ \bibinfo {pages} {1463} (\bibinfo {year} {1999})},\ \Eprint
  {http://arxiv.org/abs/hep-ph/9803479} {arXiv:hep-ph/9803479 [hep-ph]}
  \BibitemShut {NoStop}%
\bibitem [{\citenamefont {Bernreuther}(2002)}]{Bernreuther:2002uj}%
  \BibitemOpen
  \bibfield  {author} {\bibinfo {author} {\bibfnamefont {W.}~\bibnamefont
  {Bernreuther}},\ }\bibfield  {booktitle} {\emph {\bibinfo {booktitle}
  {{Workshop of the Graduate College of Elementary Particle Physics Berlin,
  Germany, April 2-5, 2001}}},\ }\href@noop {} {\bibfield  {journal} {\bibinfo
  {journal} {Lect. Notes Phys.}\ }\textbf {\bibinfo {volume} {591}},\ \bibinfo
  {pages} {237} (\bibinfo {year} {2002})},\ \bibinfo {note} {[,237(2002)]},\
  \Eprint {http://arxiv.org/abs/hep-ph/0205279} {arXiv:hep-ph/0205279 [hep-ph]}
  \BibitemShut {NoStop}%
\bibitem [{\citenamefont {Cline}(2006)}]{Cline:2006ts}%
  \BibitemOpen
  \bibfield  {author} {\bibinfo {author} {\bibfnamefont {J.~M.}\ \bibnamefont
  {Cline}},\ }in\ \href@noop {} {\emph {\bibinfo {booktitle} {{Les Houches
  Summer School - Session 86: Particle Physics and Cosmology: The Fabric of
  Spacetime Les Houches, France, July 31-August 25, 2006}}}}\ (\bibinfo {year}
  {2006})\ \Eprint {http://arxiv.org/abs/hep-ph/0609145} {arXiv:hep-ph/0609145
  [hep-ph]} \BibitemShut {NoStop}%
\bibitem [{\citenamefont {Morrissey}\ and\ \citenamefont
  {Ramsey-Musolf}(2012)}]{Morrissey:2012db}%
  \BibitemOpen
  \bibfield  {author} {\bibinfo {author} {\bibfnamefont {D.~E.}\ \bibnamefont
  {Morrissey}}\ and\ \bibinfo {author} {\bibfnamefont {M.~J.}\ \bibnamefont
  {Ramsey-Musolf}},\ }\href {\doibase 10.1088/1367-2630/14/12/125003}
  {\bibfield  {journal} {\bibinfo  {journal} {New J. Phys.}\ }\textbf {\bibinfo
  {volume} {14}},\ \bibinfo {pages} {125003} (\bibinfo {year} {2012})},\
  \Eprint {http://arxiv.org/abs/1206.2942} {arXiv:1206.2942 [hep-ph]}
  \BibitemShut {NoStop}%
\bibitem [{\citenamefont {Konstandin}(2013)}]{Konstandin:2013caa}%
  \BibitemOpen
  \bibfield  {author} {\bibinfo {author} {\bibfnamefont {T.}~\bibnamefont
  {Konstandin}},\ }\href {\doibase 10.3367/UFNe.0183.201308a.0785} {\bibfield
  {journal} {\bibinfo  {journal} {Phys. Usp.}\ }\textbf {\bibinfo {volume}
  {56}},\ \bibinfo {pages} {747} (\bibinfo {year} {2013})},\ \bibinfo {note}
  {[Usp. Fiz. Nauk183,785(2013)]},\ \Eprint {http://arxiv.org/abs/1302.6713}
  {arXiv:1302.6713 [hep-ph]} \BibitemShut {NoStop}%
\bibitem [{\citenamefont {Senaha}(2020)}]{Senaha:2020mop}%
  \BibitemOpen
  \bibfield  {author} {\bibinfo {author} {\bibfnamefont {E.}~\bibnamefont
  {Senaha}},\ }\href {\doibase 10.3390/sym12050733} {\bibfield  {journal}
  {\bibinfo  {journal} {Symmetry}\ }\textbf {\bibinfo {volume} {12}},\ \bibinfo
  {pages} {733} (\bibinfo {year} {2020})}\BibitemShut {NoStop}%
\bibitem [{\citenamefont {Dolan}\ and\ \citenamefont
  {Jackiw}(1974)}]{Dolan:1973qd}%
  \BibitemOpen
  \bibfield  {author} {\bibinfo {author} {\bibfnamefont {L.}~\bibnamefont
  {Dolan}}\ and\ \bibinfo {author} {\bibfnamefont {R.}~\bibnamefont {Jackiw}},\
  }\href {\doibase 10.1103/PhysRevD.9.3320} {\bibfield  {journal} {\bibinfo
  {journal} {Phys. Rev.}\ }\textbf {\bibinfo {volume} {D9}},\ \bibinfo {pages}
  {3320} (\bibinfo {year} {1974})}\BibitemShut {NoStop}%
\bibitem [{\citenamefont {Linde}(1980)}]{Linde:1980ts}%
  \BibitemOpen
  \bibfield  {author} {\bibinfo {author} {\bibfnamefont {A.~D.}\ \bibnamefont
  {Linde}},\ }\href {\doibase 10.1016/0370-2693(80)90769-8} {\bibfield
  {journal} {\bibinfo  {journal} {Phys. Lett.}\ }\textbf {\bibinfo {volume}
  {96B}},\ \bibinfo {pages} {289} (\bibinfo {year} {1980})}\BibitemShut
  {NoStop}%
\bibitem [{\citenamefont {Parwani}(1992)}]{Parwani:1991gq}%
  \BibitemOpen
  \bibfield  {author} {\bibinfo {author} {\bibfnamefont {R.~R.}\ \bibnamefont
  {Parwani}},\ }\href {\doibase 10.1103/PhysRevD.45.4695,
  10.1103/PhysRevD.48.5965.2} {\bibfield  {journal} {\bibinfo  {journal} {Phys.
  Rev.}\ }\textbf {\bibinfo {volume} {D45}},\ \bibinfo {pages} {4695} (\bibinfo
  {year} {1992})},\ \bibinfo {note} {[Erratum: Phys. Rev.D48,5965(1993)]},\
  \Eprint {http://arxiv.org/abs/hep-ph/9204216} {arXiv:hep-ph/9204216 [hep-ph]}
  \BibitemShut {NoStop}%
\bibitem [{\citenamefont {Carrington}(1992)}]{Carrington:1991hz}%
  \BibitemOpen
  \bibfield  {author} {\bibinfo {author} {\bibfnamefont {M.~E.}\ \bibnamefont
  {Carrington}},\ }\href {\doibase 10.1103/PhysRevD.45.2933} {\bibfield
  {journal} {\bibinfo  {journal} {Phys. Rev. D}\ }\textbf {\bibinfo {volume}
  {45}},\ \bibinfo {pages} {2933} (\bibinfo {year} {1992})}\BibitemShut
  {NoStop}%
\bibitem [{\citenamefont {Arnold}\ and\ \citenamefont
  {Espinosa}(1993)}]{Arnold:1992rz}%
  \BibitemOpen
  \bibfield  {author} {\bibinfo {author} {\bibfnamefont {P.~B.}\ \bibnamefont
  {Arnold}}\ and\ \bibinfo {author} {\bibfnamefont {O.}~\bibnamefont
  {Espinosa}},\ }\href {\doibase 10.1103/physrevd.50.6662.2,
  10.1103/PhysRevD.47.3546} {\bibfield  {journal} {\bibinfo  {journal} {Phys.
  Rev.}\ }\textbf {\bibinfo {volume} {D47}},\ \bibinfo {pages} {3546} (\bibinfo
  {year} {1993})},\ \bibinfo {note} {[Erratum: Phys. Rev.D50,6662(1994)]},\
  \Eprint {http://arxiv.org/abs/hep-ph/9212235} {arXiv:hep-ph/9212235 [hep-ph]}
  \BibitemShut {NoStop}%
\bibitem [{\citenamefont {Gould}\ and\ \citenamefont
  {Tenkanen}(2021)}]{Gould:2021oba}%
  \BibitemOpen
  \bibfield  {author} {\bibinfo {author} {\bibfnamefont {O.}~\bibnamefont
  {Gould}}\ and\ \bibinfo {author} {\bibfnamefont {T.~V.~I.}\ \bibnamefont
  {Tenkanen}},\ }\href {\doibase 10.1007/JHEP06(2021)069} {\bibfield  {journal}
  {\bibinfo  {journal} {JHEP}\ }\textbf {\bibinfo {volume} {06}},\ \bibinfo
  {pages} {069} (\bibinfo {year} {2021})},\ \Eprint
  {http://arxiv.org/abs/2104.04399} {arXiv:2104.04399 [hep-ph]} \BibitemShut
  {NoStop}%
\bibitem [{\citenamefont {Croon}\ \emph {et~al.}(2021)\citenamefont {Croon},
  \citenamefont {Gould}, \citenamefont {Schicho}, \citenamefont {Tenkanen},\
  and\ \citenamefont {White}}]{Croon:2020cgk}%
  \BibitemOpen
  \bibfield  {author} {\bibinfo {author} {\bibfnamefont {D.}~\bibnamefont
  {Croon}}, \bibinfo {author} {\bibfnamefont {O.}~\bibnamefont {Gould}},
  \bibinfo {author} {\bibfnamefont {P.}~\bibnamefont {Schicho}}, \bibinfo
  {author} {\bibfnamefont {T.~V.~I.}\ \bibnamefont {Tenkanen}}, \ and\ \bibinfo
  {author} {\bibfnamefont {G.}~\bibnamefont {White}},\ }\href {\doibase
  10.1007/JHEP04(2021)055} {\bibfield  {journal} {\bibinfo  {journal} {JHEP}\
  }\textbf {\bibinfo {volume} {04}},\ \bibinfo {pages} {055} (\bibinfo {year}
  {2021})},\ \Eprint {http://arxiv.org/abs/2009.10080} {arXiv:2009.10080
  [hep-ph]} \BibitemShut {NoStop}%
\bibitem [{\citenamefont {Schicho}\ \emph {et~al.}(2022)\citenamefont
  {Schicho}, \citenamefont {Tenkanen},\ and\ \citenamefont
  {White}}]{Schicho:2022wty}%
  \BibitemOpen
  \bibfield  {author} {\bibinfo {author} {\bibfnamefont {P.}~\bibnamefont
  {Schicho}}, \bibinfo {author} {\bibfnamefont {T.~V.~I.}\ \bibnamefont
  {Tenkanen}}, \ and\ \bibinfo {author} {\bibfnamefont {G.}~\bibnamefont
  {White}},\ }\href {\doibase 10.1007/JHEP11(2022)047} {\bibfield  {journal}
  {\bibinfo  {journal} {JHEP}\ }\textbf {\bibinfo {volume} {11}},\ \bibinfo
  {pages} {047} (\bibinfo {year} {2022})},\ \Eprint
  {http://arxiv.org/abs/2203.04284} {arXiv:2203.04284 [hep-ph]} \BibitemShut
  {NoStop}%
\bibitem [{\citenamefont {Athron}\ \emph {et~al.}(2023)\citenamefont {Athron},
  \citenamefont {Balazs}, \citenamefont {Fowlie}, \citenamefont {Morris},
  \citenamefont {White},\ and\ \citenamefont {Zhang}}]{Athron:2022jyi}%
  \BibitemOpen
  \bibfield  {author} {\bibinfo {author} {\bibfnamefont {P.}~\bibnamefont
  {Athron}}, \bibinfo {author} {\bibfnamefont {C.}~\bibnamefont {Balazs}},
  \bibinfo {author} {\bibfnamefont {A.}~\bibnamefont {Fowlie}}, \bibinfo
  {author} {\bibfnamefont {L.}~\bibnamefont {Morris}}, \bibinfo {author}
  {\bibfnamefont {G.}~\bibnamefont {White}}, \ and\ \bibinfo {author}
  {\bibfnamefont {Y.}~\bibnamefont {Zhang}},\ }\href {\doibase
  10.1007/JHEP01(2023)050} {\bibfield  {journal} {\bibinfo  {journal} {JHEP}\
  }\textbf {\bibinfo {volume} {01}},\ \bibinfo {pages} {050} (\bibinfo {year}
  {2023})},\ \Eprint {http://arxiv.org/abs/2208.01319} {arXiv:2208.01319
  [hep-ph]} \BibitemShut {NoStop}%
\bibitem [{\citenamefont {Coleman}\ and\ \citenamefont
  {Weinberg}(1973)}]{Coleman:1973jx}%
  \BibitemOpen
  \bibfield  {author} {\bibinfo {author} {\bibfnamefont {S.~R.}\ \bibnamefont
  {Coleman}}\ and\ \bibinfo {author} {\bibfnamefont {E.~J.}\ \bibnamefont
  {Weinberg}},\ }\href {\doibase 10.1103/PhysRevD.7.1888} {\bibfield  {journal}
  {\bibinfo  {journal} {Phys. Rev. D}\ }\textbf {\bibinfo {volume} {7}},\
  \bibinfo {pages} {1888} (\bibinfo {year} {1973})}\BibitemShut {NoStop}%
\bibitem [{\citenamefont {Kastening}(1992)}]{Kastening:1991gv}%
  \BibitemOpen
  \bibfield  {author} {\bibinfo {author} {\bibfnamefont {B.~M.}\ \bibnamefont
  {Kastening}},\ }\href {\doibase 10.1016/0370-2693(92)90021-U} {\bibfield
  {journal} {\bibinfo  {journal} {Phys. Lett. B}\ }\textbf {\bibinfo {volume}
  {283}},\ \bibinfo {pages} {287} (\bibinfo {year} {1992})}\BibitemShut
  {NoStop}%
\bibitem [{\citenamefont {Bando}\ \emph
  {et~al.}(1993{\natexlab{a}})\citenamefont {Bando}, \citenamefont {Kugo},
  \citenamefont {Maekawa},\ and\ \citenamefont {Nakano}}]{Bando:1992np}%
  \BibitemOpen
  \bibfield  {author} {\bibinfo {author} {\bibfnamefont {M.}~\bibnamefont
  {Bando}}, \bibinfo {author} {\bibfnamefont {T.}~\bibnamefont {Kugo}},
  \bibinfo {author} {\bibfnamefont {N.}~\bibnamefont {Maekawa}}, \ and\
  \bibinfo {author} {\bibfnamefont {H.}~\bibnamefont {Nakano}},\ }\href
  {\doibase 10.1016/0370-2693(93)90725-W} {\bibfield  {journal} {\bibinfo
  {journal} {Phys. Lett. B}\ }\textbf {\bibinfo {volume} {301}},\ \bibinfo
  {pages} {83} (\bibinfo {year} {1993}{\natexlab{a}})},\ \Eprint
  {http://arxiv.org/abs/hep-ph/9210228} {arXiv:hep-ph/9210228} \BibitemShut
  {NoStop}%
\bibitem [{\citenamefont {Bando}\ \emph
  {et~al.}(1993{\natexlab{b}})\citenamefont {Bando}, \citenamefont {Kugo},
  \citenamefont {Maekawa},\ and\ \citenamefont {Nakano}}]{Bando:1992wy}%
  \BibitemOpen
  \bibfield  {author} {\bibinfo {author} {\bibfnamefont {M.}~\bibnamefont
  {Bando}}, \bibinfo {author} {\bibfnamefont {T.}~\bibnamefont {Kugo}},
  \bibinfo {author} {\bibfnamefont {N.}~\bibnamefont {Maekawa}}, \ and\
  \bibinfo {author} {\bibfnamefont {H.}~\bibnamefont {Nakano}},\ }\href
  {\doibase 10.1143/PTP.90.405} {\bibfield  {journal} {\bibinfo  {journal}
  {Prog. Theor. Phys.}\ }\textbf {\bibinfo {volume} {90}},\ \bibinfo {pages}
  {405} (\bibinfo {year} {1993}{\natexlab{b}})},\ \Eprint
  {http://arxiv.org/abs/hep-ph/9210229} {arXiv:hep-ph/9210229} \BibitemShut
  {NoStop}%
\bibitem [{\citenamefont {Ford}\ \emph {et~al.}(1993)\citenamefont {Ford},
  \citenamefont {Jones}, \citenamefont {Stephenson},\ and\ \citenamefont
  {Einhorn}}]{Ford:1992mv}%
  \BibitemOpen
  \bibfield  {author} {\bibinfo {author} {\bibfnamefont {C.}~\bibnamefont
  {Ford}}, \bibinfo {author} {\bibfnamefont {D.~R.~T.}\ \bibnamefont {Jones}},
  \bibinfo {author} {\bibfnamefont {P.~W.}\ \bibnamefont {Stephenson}}, \ and\
  \bibinfo {author} {\bibfnamefont {M.~B.}\ \bibnamefont {Einhorn}},\ }\href
  {\doibase 10.1016/0550-3213(93)90206-5} {\bibfield  {journal} {\bibinfo
  {journal} {Nucl. Phys. B}\ }\textbf {\bibinfo {volume} {395}},\ \bibinfo
  {pages} {17} (\bibinfo {year} {1993})},\ \Eprint
  {http://arxiv.org/abs/hep-lat/9210033} {arXiv:hep-lat/9210033} \BibitemShut
  {NoStop}%
\bibitem [{\citenamefont {Machacek}\ and\ \citenamefont
  {Vaughn}(1983)}]{Machacek:1983tz}%
  \BibitemOpen
  \bibfield  {author} {\bibinfo {author} {\bibfnamefont {M.~E.}\ \bibnamefont
  {Machacek}}\ and\ \bibinfo {author} {\bibfnamefont {M.~T.}\ \bibnamefont
  {Vaughn}},\ }\href {\doibase 10.1016/0550-3213(83)90610-7} {\bibfield
  {journal} {\bibinfo  {journal} {Nucl. Phys. B}\ }\textbf {\bibinfo {volume}
  {222}},\ \bibinfo {pages} {83} (\bibinfo {year} {1983})}\BibitemShut
  {NoStop}%
\bibitem [{\citenamefont {Machacek}\ and\ \citenamefont
  {Vaughn}(1984)}]{Machacek:1983fi}%
  \BibitemOpen
  \bibfield  {author} {\bibinfo {author} {\bibfnamefont {M.~E.}\ \bibnamefont
  {Machacek}}\ and\ \bibinfo {author} {\bibfnamefont {M.~T.}\ \bibnamefont
  {Vaughn}},\ }\href {\doibase 10.1016/0550-3213(84)90533-9} {\bibfield
  {journal} {\bibinfo  {journal} {Nucl. Phys. B}\ }\textbf {\bibinfo {volume}
  {236}},\ \bibinfo {pages} {221} (\bibinfo {year} {1984})}\BibitemShut
  {NoStop}%
\bibitem [{\citenamefont {Machacek}\ and\ \citenamefont
  {Vaughn}(1985)}]{Machacek:1984zw}%
  \BibitemOpen
  \bibfield  {author} {\bibinfo {author} {\bibfnamefont {M.~E.}\ \bibnamefont
  {Machacek}}\ and\ \bibinfo {author} {\bibfnamefont {M.~T.}\ \bibnamefont
  {Vaughn}},\ }\href {\doibase 10.1016/0550-3213(85)90040-9} {\bibfield
  {journal} {\bibinfo  {journal} {Nucl. Phys. B}\ }\textbf {\bibinfo {volume}
  {249}},\ \bibinfo {pages} {70} (\bibinfo {year} {1985})}\BibitemShut
  {NoStop}%
\bibitem [{\citenamefont {Luo}\ and\ \citenamefont {Xiao}(2003)}]{Luo:2002ey}%
  \BibitemOpen
  \bibfield  {author} {\bibinfo {author} {\bibfnamefont {M.-x.}\ \bibnamefont
  {Luo}}\ and\ \bibinfo {author} {\bibfnamefont {Y.}~\bibnamefont {Xiao}},\
  }\href {\doibase 10.1103/PhysRevLett.90.011601} {\bibfield  {journal}
  {\bibinfo  {journal} {Phys. Rev. Lett.}\ }\textbf {\bibinfo {volume} {90}},\
  \bibinfo {pages} {011601} (\bibinfo {year} {2003})},\ \Eprint
  {http://arxiv.org/abs/hep-ph/0207271} {arXiv:hep-ph/0207271} \BibitemShut
  {NoStop}%
\bibitem [{\citenamefont {Luo}\ \emph {et~al.}(2003)\citenamefont {Luo},
  \citenamefont {Wang},\ and\ \citenamefont {Xiao}}]{Luo:2002ti}%
  \BibitemOpen
  \bibfield  {author} {\bibinfo {author} {\bibfnamefont {M.-x.}\ \bibnamefont
  {Luo}}, \bibinfo {author} {\bibfnamefont {H.-w.}\ \bibnamefont {Wang}}, \
  and\ \bibinfo {author} {\bibfnamefont {Y.}~\bibnamefont {Xiao}},\ }\href
  {\doibase 10.1103/PhysRevD.67.065019} {\bibfield  {journal} {\bibinfo
  {journal} {Phys. Rev. D}\ }\textbf {\bibinfo {volume} {67}},\ \bibinfo
  {pages} {065019} (\bibinfo {year} {2003})},\ \Eprint
  {http://arxiv.org/abs/hep-ph/0211440} {arXiv:hep-ph/0211440} \BibitemShut
  {NoStop}%
\bibitem [{\citenamefont {Funakubo}\ and\ \citenamefont
  {Senaha}(2023)}]{Funakubo:2023cyv}%
  \BibitemOpen
  \bibfield  {author} {\bibinfo {author} {\bibfnamefont {K.}~\bibnamefont
  {Funakubo}}\ and\ \bibinfo {author} {\bibfnamefont {E.}~\bibnamefont
  {Senaha}},\ }\href@noop {} {\  (\bibinfo {year} {2023})},\ \Eprint
  {http://arxiv.org/abs/2307.02153} {arXiv:2307.02153 [hep-ph]} \BibitemShut
  {NoStop}%
\bibitem [{\citenamefont {'t~Hooft}\ and\ \citenamefont
  {Veltman}(1972)}]{tHooft:1972tcz}%
  \BibitemOpen
  \bibfield  {author} {\bibinfo {author} {\bibfnamefont {G.}~\bibnamefont
  {'t~Hooft}}\ and\ \bibinfo {author} {\bibfnamefont {M.~J.~G.}\ \bibnamefont
  {Veltman}},\ }\href {\doibase 10.1016/0550-3213(72)90279-9} {\bibfield
  {journal} {\bibinfo  {journal} {Nucl. Phys. B}\ }\textbf {\bibinfo {volume}
  {44}},\ \bibinfo {pages} {189} (\bibinfo {year} {1972})}\BibitemShut
  {NoStop}%
\bibitem [{\citenamefont {Kneur}\ and\ \citenamefont
  {Pinto}(2016)}]{Kneur:2015uha}%
  \BibitemOpen
  \bibfield  {author} {\bibinfo {author} {\bibfnamefont {J.~L.}\ \bibnamefont
  {Kneur}}\ and\ \bibinfo {author} {\bibfnamefont {M.~B.}\ \bibnamefont
  {Pinto}},\ }\href {\doibase 10.1103/PhysRevLett.116.031601} {\bibfield
  {journal} {\bibinfo  {journal} {Phys. Rev. Lett.}\ }\textbf {\bibinfo
  {volume} {116}},\ \bibinfo {pages} {031601} (\bibinfo {year} {2016})},\
  \Eprint {http://arxiv.org/abs/1507.03508} {arXiv:1507.03508 [hep-ph]}
  \BibitemShut {NoStop}%
\bibitem [{\citenamefont {Kneur}\ and\ \citenamefont
  {Pinto}()}]{Kneur:2015moa}%
  \BibitemOpen
  \bibfield  {author} {\bibinfo {author} {\bibfnamefont {J.~L.}\ \bibnamefont
  {Kneur}}\ and\ \bibinfo {author} {\bibfnamefont {M.~B.}\ \bibnamefont
  {Pinto}},\ }\href {\doibase 10.1103/PhysRevD.92.116008} {\bibfield  {journal}
  {\bibinfo  {journal} {Phys. Rev. D}\ }\textbf {\bibinfo {volume} {92}},\
  \bibinfo {pages} {116008}},\ \Eprint {http://arxiv.org/abs/1508.02610}
  {arXiv:1508.02610 [hep-ph]} \BibitemShut {NoStop}%
\bibitem [{\citenamefont {Karsch}\ \emph {et~al.}(1997)\citenamefont {Karsch},
  \citenamefont {Patkos},\ and\ \citenamefont {Petreczky}}]{Karsch:1997gj}%
  \BibitemOpen
  \bibfield  {author} {\bibinfo {author} {\bibfnamefont {F.}~\bibnamefont
  {Karsch}}, \bibinfo {author} {\bibfnamefont {A.}~\bibnamefont {Patkos}}, \
  and\ \bibinfo {author} {\bibfnamefont {P.}~\bibnamefont {Petreczky}},\ }\href
  {\doibase 10.1016/S0370-2693(97)00392-4} {\bibfield  {journal} {\bibinfo
  {journal} {Phys. Lett. B}\ }\textbf {\bibinfo {volume} {401}},\ \bibinfo
  {pages} {69} (\bibinfo {year} {1997})},\ \Eprint
  {http://arxiv.org/abs/hep-ph/9702376} {arXiv:hep-ph/9702376} \BibitemShut
  {NoStop}%
\bibitem [{\citenamefont {Laine}\ \emph {et~al.}(2017)\citenamefont {Laine},
  \citenamefont {Meyer},\ and\ \citenamefont {Nardini}}]{Laine:2017hdk}%
  \BibitemOpen
  \bibfield  {author} {\bibinfo {author} {\bibfnamefont {M.}~\bibnamefont
  {Laine}}, \bibinfo {author} {\bibfnamefont {M.}~\bibnamefont {Meyer}}, \ and\
  \bibinfo {author} {\bibfnamefont {G.}~\bibnamefont {Nardini}},\ }\href
  {\doibase 10.1016/j.nuclphysb.2017.04.023} {\bibfield  {journal} {\bibinfo
  {journal} {Nucl. Phys.}\ }\textbf {\bibinfo {volume} {B920}},\ \bibinfo
  {pages} {565} (\bibinfo {year} {2017})},\ \Eprint
  {http://arxiv.org/abs/1702.07479} {arXiv:1702.07479 [hep-ph]} \BibitemShut
  {NoStop}%
\bibitem [{\citenamefont {Banerjee}\ and\ \citenamefont
  {Mallik}(1991)}]{Banerjee:1991fu}%
  \BibitemOpen
  \bibfield  {author} {\bibinfo {author} {\bibfnamefont {N.}~\bibnamefont
  {Banerjee}}\ and\ \bibinfo {author} {\bibfnamefont {S.}~\bibnamefont
  {Mallik}},\ }\href {\doibase 10.1103/PhysRevD.43.3368} {\bibfield  {journal}
  {\bibinfo  {journal} {Phys. Rev. D}\ }\textbf {\bibinfo {volume} {43}},\
  \bibinfo {pages} {3368} (\bibinfo {year} {1991})}\BibitemShut {NoStop}%
\bibitem [{\citenamefont {Chiku}\ and\ \citenamefont
  {Hatsuda}(1998)}]{Chiku:1998kd}%
  \BibitemOpen
  \bibfield  {author} {\bibinfo {author} {\bibfnamefont {S.}~\bibnamefont
  {Chiku}}\ and\ \bibinfo {author} {\bibfnamefont {T.}~\bibnamefont
  {Hatsuda}},\ }\href {\doibase 10.1103/PhysRevD.58.076001} {\bibfield
  {journal} {\bibinfo  {journal} {Phys. Rev. D}\ }\textbf {\bibinfo {volume}
  {58}},\ \bibinfo {pages} {076001} (\bibinfo {year} {1998})},\ \Eprint
  {http://arxiv.org/abs/hep-ph/9803226} {arXiv:hep-ph/9803226} \BibitemShut
  {NoStop}%
\bibitem [{\citenamefont {Funakubo}\ and\ \citenamefont
  {Senaha}(2013)}]{Funakubo:2012qc}%
  \BibitemOpen
  \bibfield  {author} {\bibinfo {author} {\bibfnamefont {K.}~\bibnamefont
  {Funakubo}}\ and\ \bibinfo {author} {\bibfnamefont {E.}~\bibnamefont
  {Senaha}},\ }\href {\doibase 10.1103/PhysRevD.87.054003} {\bibfield
  {journal} {\bibinfo  {journal} {Phys. Rev. D}\ }\textbf {\bibinfo {volume}
  {87}},\ \bibinfo {pages} {054003} (\bibinfo {year} {2013})},\ \Eprint
  {http://arxiv.org/abs/1210.1737} {arXiv:1210.1737 [hep-ph]} \BibitemShut
  {NoStop}%
\bibitem [{\citenamefont {Funakubo}\ and\ \citenamefont {Senaha}()}]{FS}%
  \BibitemOpen
  \bibfield  {author} {\bibinfo {author} {\bibfnamefont {K.}~\bibnamefont
  {Funakubo}}\ and\ \bibinfo {author} {\bibfnamefont {E.}~\bibnamefont
  {Senaha}},\ }\href@noop {} {\ }\bibinfo {note} {In preparation}\BibitemShut
  {NoStop}%
\bibitem [{\citenamefont {Boyd}\ \emph {et~al.}(1993)\citenamefont {Boyd},
  \citenamefont {Brahm},\ and\ \citenamefont {Hsu}}]{Boyd:1993tz}%
  \BibitemOpen
  \bibfield  {author} {\bibinfo {author} {\bibfnamefont {C.~G.}\ \bibnamefont
  {Boyd}}, \bibinfo {author} {\bibfnamefont {D.~E.}\ \bibnamefont {Brahm}}, \
  and\ \bibinfo {author} {\bibfnamefont {S.~D.~H.}\ \bibnamefont {Hsu}},\
  }\href {\doibase 10.1103/PhysRevD.48.4963} {\bibfield  {journal} {\bibinfo
  {journal} {Phys. Rev. D}\ }\textbf {\bibinfo {volume} {48}},\ \bibinfo
  {pages} {4963} (\bibinfo {year} {1993})},\ \Eprint
  {http://arxiv.org/abs/hep-ph/9304254} {arXiv:hep-ph/9304254} \BibitemShut
  {NoStop}%
\bibitem [{\citenamefont {Curtin}\ \emph {et~al.}(2018)\citenamefont {Curtin},
  \citenamefont {Meade},\ and\ \citenamefont {Ramani}}]{Curtin:2016urg}%
  \BibitemOpen
  \bibfield  {author} {\bibinfo {author} {\bibfnamefont {D.}~\bibnamefont
  {Curtin}}, \bibinfo {author} {\bibfnamefont {P.}~\bibnamefont {Meade}}, \
  and\ \bibinfo {author} {\bibfnamefont {H.}~\bibnamefont {Ramani}},\ }\href
  {\doibase 10.1140/epjc/s10052-018-6268-0} {\bibfield  {journal} {\bibinfo
  {journal} {Eur. Phys. J. C}\ }\textbf {\bibinfo {volume} {78}},\ \bibinfo
  {pages} {787} (\bibinfo {year} {2018})},\ \Eprint
  {http://arxiv.org/abs/1612.00466} {arXiv:1612.00466 [hep-ph]} \BibitemShut
  {NoStop}%
\end{thebibliography}%
%

\end{document}